\documentclass[lettersize,journal]{IEEEtran}  
\usepackage{amsmath, amsfonts}
\usepackage{amsmath}  
\usepackage{amsthm}
\usepackage{amssymb}
\usepackage{array}
\usepackage[caption=false,font=normalsize,labelfont=sf,textfont=sf]{subfig}
\usepackage{textcomp}
\usepackage{color}
\usepackage{stfloats}
\usepackage{url}
\usepackage{verbatim}                   
\usepackage{xcolor}
\usepackage{graphicx}
\usepackage[colorlinks, citecolor=black,linkcolor=black ]{hyperref}      
\usepackage{cite}
\usepackage[ruled,linesnumbered]{algorithm2e}
\usepackage{booktabs}       
\usepackage{verbatim}      
\usepackage{bbding}
\usepackage{threeparttable}
\usepackage{multirow}
\usepackage{tabularx}
\usepackage{makecell}
\usepackage{xspace}
\usepackage{textcomp}
\makeatletter

\newcommand{\Rmnum}[1]{\expandafter@slowromancap\romannumeral #1@}   
\makeatother
\begin{document}
	\title{Interference Management for Integrated Sensing and Communication Systems: A Survey}
\author{\IEEEauthorblockN{Yangyang Niu, \emph{Student Member, IEEE}, Zhiqing Wei, \emph{Member, IEEE}, Lin Wang, \emph{Student Member, IEEE},\\
		 Huici Wu, \emph{Member, IEEE}, and Zhiyong Feng, \emph{Senior Member, IEEE}}

\thanks{
		
Yangyang Niu, Zhiqing Wei, Lin Wang, Zhiyong Feng are with the Key Laboratory of Universal Wireless Communications, Ministry of Education, School of Information and Communication Engineering, Beijing University of Posts and Telecommunications, Beijing 100876, China (e-mail: niuyy@bupt.edu.cn, weizhiqing@bupt.edu.cn, wlwl@bupt.edu.cn, fengzy@bupt.edu.cn).

Huici Wu is with the National Engineering Lab for Mobile Network Technologies, Beijing University of Posts and Telecommunications, Beijing 100876, China, and also with Peng Cheng Laboratory, Shenzhen 518066, China (e-mail: dailywu@bupt.edu.cn).
	  }
}

\maketitle
\thispagestyle{empty}
\pagestyle{empty}	
\begin{abstract} \label{abstract}
	Emerging applications such as autonomous driving and Internet of things (IoT) services put forward the demand for simutaneous sensing and communication functions in the same system.
	Integrated sensing and communication (ISAC) has the potential to meet the demands of ubiquitous communication and high-precision sensing due to the advantages of spectrum and hardware resource sharing, as well as the mutual enhancement of sensing and communication.  
	However, ISAC system faces severe interference requiring effective interference suppression, avoidance, and exploitation techniques. 
	This article provides a comprehensive survey on the interference management techniques in ISAC systems, involving network architecture, system design, signal processing, and resource allocation. 
	We first review the channel modeling and performance metrics of the ISAC system. 
	Then, the methods for managing self-interference (SI), mutual interference (MI), and clutter in a single base station (BS) system are summarized, including interference suppression, interference avoidance and interference exploitation methods.  
	Furthermore, cooperative interference management methods are studied to address the cross-link interference (CLI) in a coordinated multi-point ISAC (CoMP-ISAC) system. 
	Finally, future trends are revealed. 
	This article may provide a reference for the study of interference management in ISAC systems. 
	
\end{abstract}


\section{Introduction}	
\subsection{Background and Motivation}	
\begin{figure*}
	\centering
	\includegraphics[width=1\linewidth]{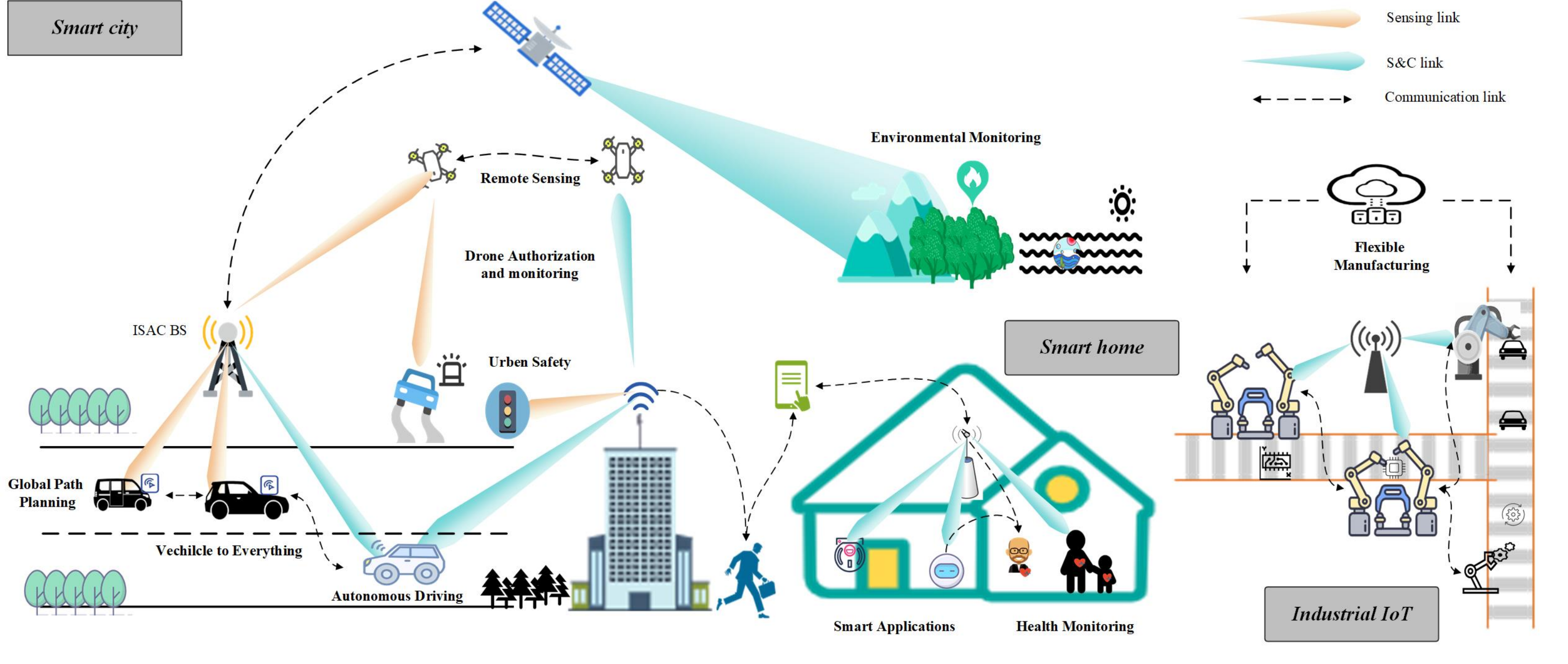}
	\caption{The scenarios and applications of ISAC systems.}
	\label{scene}
\end{figure*}
	With the rapid development of big data technology and artificial intelligence (AI), Internet of things (IoT) is experiencing continuous enhancement \cite{9606831}. Emerging AI services, immersive experiences, and digital twin services are continuously evolving, especially in the scenarios such as smart city, smart home, and industrial IoT \cite{cui2023physical}. These applications not only demand dual functionalities of sensing and communication (S\&C) but also pose high requirements for end-to-end information processing capabilities, involving multiple stages such as data transmission, sensing, prediction, control, and decision-making \cite{tallat2023navigating}. 

	Integrated sensing and communication (ISAC) refers to the design paradigm in which S\&C functionalities are integrated to efficiently utilize mutual benefits by sharing hardware frameworks, spectrum resources, and signal processing \cite{ouyang2023mimo}. 
	ISAC supports ubiquitous communication and precise sensing throughout various applications, as illustrated in Fig.  \ref{scene}. 
	In recent years, ISAC has attracted widespread attention. 
	For instance, Federal Communications Commission (FCC) and National Telecommunications and Information Administration (NTIA) in United States have proposed sharing 150 MHz of spectrum allocated for sensing applications in the 3.5 GHz band with communication applications \cite{locke2010assessment}.
	In addition, International Telecommunication Union (ITU) officially designated ISAC as one of the six key scenarios for the sixth-generation (6G) of mobile communication system in 2023, demonstrating its significant potential in the era of ubiquitous intelligence \cite{ITU-R-IMT-2030}. 
	The advancements in mobile communication such as millimeter-wave (mmWave), terahertz (THz), massive multiple-input multiple-output (MIMO), reconfigurable intelligent surface (RIS), and advanced signal processing algorithms, have promoted the realization of ISAC \cite{ 9528013, 10184278, 9309152}.	
	Nevertheless, compared to communication-only and sensing-only systems, ISAC systems suffer from both interference from sensing and communication and more complex inter-base station (BS) crosstalk, which specifically include the following four types.
	
	\subsubsection{Self-Interference (SI) }
	ISAC offers more flexibility than traditional frequency-division S\&C schemes due to the shared spectrum and air-interface resources \cite{10418473}. 
	Nerveless, echo signals face significant SI as it arrives at the sensing receiver before the completion of ISAC signal transmission.
	
 	\subsubsection{Mutual-Interference (MI) }
	In the downlink transmission, the reliable transmission of communication signals is influenced by sensing signals with greater signal strength. 
	In the uplink transmission, MI between uplink communication  signals and echo signals bring challenge for target recognition and localization \cite{9800940, 10036139}. 
			
	\subsubsection{Clutter }
	In the complex scenarios such as industrial environments and urban areas, ISAC system faces challenges of stationary clutter and dynamic clutter from environment and target \cite{8827589}. 		
			
	\subsubsection{Cross-Link Interference (CLI)}
	To fully leverage the diversity gained from widely distributed sensing nodes, constructing a coordinated multi-point ISAC (CoMP-ISAC) system is an inevitable trend \cite{10273396}. However, coordinating signal transmission and sensing signal processing over a large area presents complex CLI among multiple BSs \cite{9139384, jiang2023integrated }.			

	To tackle the above challenges, this article provides an in-depth review on interference suppression, interference avoidance, and interference exploitation.
	
\subsection{Existing Surveys and Tutorials}	
	The existing survey and tutorial articles of ISAC mainly focus on system design and signal processing, where interference managements are rarely mentioned.
	For instance, Zhang \textit{et al.} comprehensively reviewed the latest advancements in joint communication and radar (JCR) systems from a signal processing perspective, emphasizing advanced signal processing techniques to suppress interference and enhance integrated gain\cite{9540344}. 
	Wei \textit{et al.} provided a detailed review of signal processing methods in ISAC systems from signal design, signal processing, and signal optimization, highlighting the advantages of collaborative signal processing \cite{10012421}. 
	In addition, Fang \textit{et al.} focused on the application of MIMO in ISAC, revealing that MIMO can construct favorable channels for signal transmission in the spatial domain, which is conducive to large-scale deployment due to its low cost and effective interference suppression \cite{9971740}. 
	Moreover, Liu \textit{et al.} carried out a comprehensive review on the performance trade-off, signal processing, and network architecture of ISAC systems, while also addressing the limitations of interference management in enhancing the performance of ISAC systems \cite{9737357}.

\begin{table*}[t]
	\centering
	\setlength{\extrarowheight}{2pt}
	\begin{threeparttable}
	\centering
	\caption { Review articles covering interference management in ISAC}
	\label{refs}
	\begin{tabular}{|c|c|c|c|c|c|} 
		\hline
	\textbf{Reference}&	\textbf{Year} & \textbf{ Interference Type}   &  \textbf{Interference Management Method} &  \textbf{Transmission Direction}	 & \textbf{CoMP-ISAC} \\ \hline			
	\cite{Li2020ATO} &   2020  &  2      & 2, 3  & 1      &             \\ \hline
	\cite{8972666}  &  2020    &   2     & 1, 2  & 1      &               \\ \hline
	\cite{9139384}  & 2020     &  2,4  & 1, 2, 3 &  1, 2  & \checkmark      \\ \hline
	\cite{9540344} & 2021      &  2      & 2     &   1    &              \\ \hline
	\cite{9393464} & 2021      &  2      & 1, 2  &   1    &              \\ \hline
	\cite{9737357} & 2022      & 1, 2, 3 & 1, 2  &  1, 2   &   \checkmark   \\ \hline
	\cite{9705498} & 2022      &  2      &1, 2   &  1, 2   &              \\ \hline
	\cite{9971740} &   2023    &   2, 3  & 1, 2  &  1     &   \checkmark   \\ \hline
	\cite{10012421} &  2023   &   1, 2   & 1, 2 &   1      &                \\ \hline
        This survey &  2024  & 1, 2, 3, 4 & 1, 2, 3   & 1,2 &   \checkmark \\ \hline
		
	\end{tabular}
\begin{tablenotes}
	\item Notes: The meaning of the numbers in the table is as follows.
	\item 	Types of interference: 1 SI, 2 MI, 3 Clutter, 4 CLI.
	\item  Interference management methods: 1 Interference cancellation, 2 Interference avoidance, 3 Interference exploitation. 
	\item Transmission direction: 1 Downlink, 2 Uplink.
\end{tablenotes}
	\end{threeparttable}
\end{table*}

Although existing reviews of ISAC offer valuable insights into system design and signal processing, they still have the following limitations. 
A comprehensive review on the interference management for different types of interference is lacking. 
Furthermore, the interference management involved in CoMP-ISAC system has not been reviewed.  

\subsection{Contributions and Structure} 
In this article, we provide a comprehensive survey on interference management in ISAC systems. Specifically, we review SI, MI, clutter, and CLI management methods from suppression, avoidance, and exploitation perspectives. 
The main differences between prior works and this article are listed in Table \ref{refs}.
In this regard, the main contributions of this article are summarized as follows. 

\subsubsection{SI Management } 
The methods for interference cancellation and avoidance are reviewed in the digital, analog, and propagation domains.
\subsubsection{MI Management } 
The methods for managing MI, including interference suppression, avoidance, and exploitation, are summarized separately for the uplink and downlink signal transmission.
\subsubsection{Clutter Management }  
Interference management methods are summarized from BS operational modes, uplink assistance, and network resource allocation perspectives.
\subsubsection{CLI Management } 
Interference management methods are summarized from the perspectives of BS operational modes, uplink assistance, and network resource allocation.

\begin{figure}
	\centering
	\includegraphics[width=1\linewidth]{ 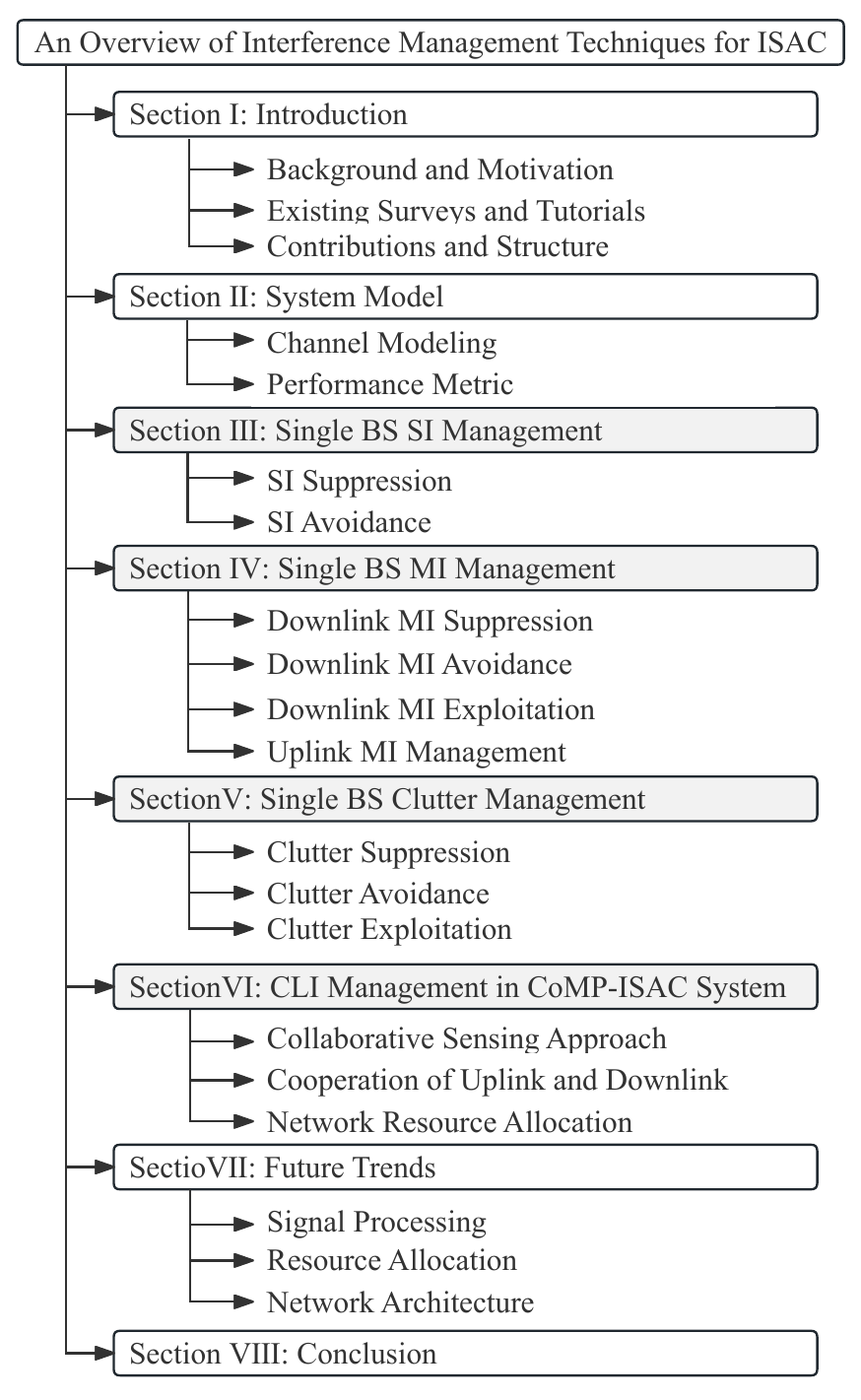}
	\caption{The organization of this article.}
	\label{structure}
\end{figure}

The structure of this article is illustrated in Fig. \ref{structure}. 
The abbreviations commonly used in this article are summarized in Table \ref{acronyms}.

\begin{table*}[]
	\setlength{\extrarowheight}{4pt}
	\centering
	\caption {List of abbreviations}
	\label{acronyms}
	\begin{tabular}{|l|l|l|l|l|p{4.3cm}|} 
		\hline
		\textbf{Abbr.}  &  \textbf{Definitions}  & \textbf{Abbr.}   &  \textbf{Definitions}& \textbf{Abbr.}  &   \textbf{Definitions}  \\ \hline	

BS  & Base station & UE & User equipment & S\&C  & Sensing and communication  \\ \hline
6G & Sixth-generation &SINR & Signal-to-interference-plus-noise ratio& SCNR & Signal-to-clutter-plus-noise-ratio  \\ \hline
SI & Self-interference & MI & Mutual-interference & CLI  & Cross-link interference  \\ \hline
MIMO  &Multiple-input multiple-output & CoMP & Coordinated multi-point & SIC & Successive interference cancellation  \\ \hline 
MUI  & Multi-user interference & NOMA & Non-orthogonal
multiple access & CIR  & Channel impulse response \\ \hline
CSI  & Channel state information & AoA &Angle-of-arrival & AoD  & Angle-of-departure \\ \hline 
SIR & Signal-to-interference Ratio & ALMS  & Analog least mean-square         &  RF &  Radio Frequency   \\ \hline
EE  &   Energy-efficiency      & NR & New radio & MUSIC  & Multiple signal classification \\ \hline
MA & Multiple access  & RIS & Reconfigurable intelligent surface & RHS  & Reconfigurable holographic surface  \\ \hline
TDMA & Time division multiple access & FDMA & Frequency division multiple access & SDMA & Spatial division multiple access  \\ \hline 
C-RAN & Cloud-radio access network & LFM & Linear frequency modulation & OFDM & Orthogonal frequency division multiplexing  \\ \hline
RSMA & Rate-splitting multiple access & DFRC &Dual-function radar-communication & JCR &  Joint communication and radar  \\ \hline   
CRLB & Cram\'{e}r-Rao lower bound  & KF &Kalman filter  & LS  & Least-squares  \\ \hline
SDN &  Software-defined networking & MCL & Mutual coupling loss   & PAPR & Peak to average power ratio  \\ \hline
CI  &Constructive interference & ISR & Interference suppression ratio & CCM  & Clutter covariance matrix \\ \hline
STAP & Space-time adaptive processing & SLAM & Simultaneous localization and mapping  & CPI  &Coherent processing interval  \\ \hline
TO &Timing offset& CFO & Carrier frequency offset & SER  & Symbol error rate \\ \hline
LoS  & Line-of-sight  & NLoS & Non-line-of-sight  & UAV & Unmanned arial vehicle \\ \hline	
ISI  & Inter-symbol interference & ULA  & Uniform linear array  & UPA   &  Uniform planar array \\ \hline	
TMT & Target monitoring terminal  &  MTI & Moving target indication   & MTD  &  Moving target detection \\ \hline	
	\end{tabular}
\end{table*}

\section{System Model}
Channel modeling serves as the foundation for understanding and predicting the behavior of wireless signals during propagation. 
Accurate channel modeling facilitates the identification and comprehension of interference sources within ISAC systems, thereby establishing a foundation for interference management. 
Furthermore, effective interference management can enhance system performance and reliability by refining the quality of channel modeling.
In addition, the performance metrics are crucial factors for measuring and evaluating the design, implementation, and optimization of ISAC systems.
Therefore, this section reviews the ISAC channel modeling and performance metrics of ISAC systems.

\subsection{Channel Modeling}

\subsubsection{Related Works}
In general, ISAC channel modeling can be divided into sensing channel modeling and communication channel modeling. 
The purpose of communication channel modeling is to enhance the accuracy of signal transmission, thus disregarding scattering in intermediate processes.
In contrast, the sensing relies on receiving echo signals reflected from the target for detection and parameter estimation. 
Nevertheless, S\&C signal transmission follows the electromagnetic propagation mechanism, offering the possibility of a unified channel modeling format. 
To accurately simulate and evaluate the performance of ISAC systems, Liu \textit{et al.} emphasized the importance of considering shared scatterers and proposed a cluster-based stochastic channel model to extend and complement existing MIMO channel models \cite{10334037}. 
The proposed cluster-based channel model is widely adopted in the scenarios incorporating stacked or extended antenna configurations because of its robustness and efficiency.
Moreover, Yang \textit{et al.} proposed an ISAC channel modeling considering both forward scattering and backward scattering components on behalf of non-line-of-sight (NLoS) except line-of-sight (LoS) \cite{10078840}. 

\subsubsection{Sensing Channel Modeling}
Along with the above researches, this subsection provides a concise form of ISAC channel modeling including sensing channel modeling and communication channel modeling. 
\begin{figure}[t]
	\centering
	\includegraphics[width=0.8\linewidth]{ 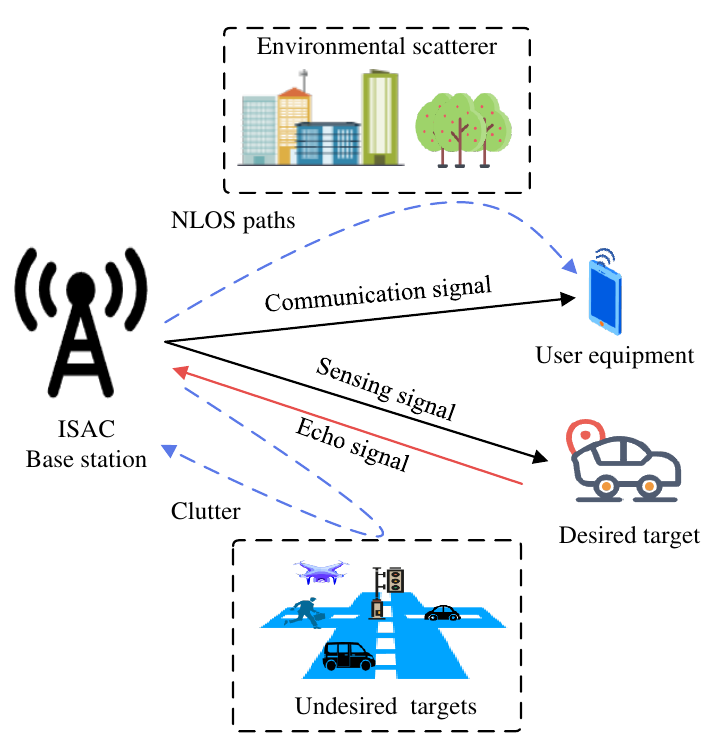}
	\caption{ISAC channel modeling \cite{10334037}.}
	\label{channel modeling}
\end{figure}

Without loss of generality, Fig. \ref{channel modeling} illustrates a broadband downlink transmission in an ISAC system consists of a BS, a user equipment (UE), and a desired sensing target, where the ISAC BS is equipped with a two-dimensional uniform planar array (UPA) placed on the $x,y$ plane.  
The size of UPA is $P\times Q$ with uniform element intervals, denoted as $d_a$. 
The index $(p,q)$ expresses antenna element at the $p$th row and $q$th column.
The phase difference between the $(p,q)$th element and the reference element, namely the (0,0)th antenna element, is given by
\begin{equation}   
	{{\left[ \mathbf{a}\left( \theta ,\phi  \right) \right]}_{p,q}}=\exp \left[ -j\frac{2\pi }{\lambda }{{d}_{a}}\left( p\cos \theta \sin \phi +q\sin \theta \sin \phi  \right) \right],
\end{equation}
where $\theta$ and $\phi $ are the azimuth and elevation angles respectively, $\lambda=c/f_c$ is the wavelength, $f_c$ is the carrier frequency, and $c$ is the speed of light.
The steering matrix $\mathbf{A}(\theta, \phi)$ can be denoted as the product of the steering vectors of transmit array and receive array as follows.
\begin{equation}
	\mathbf{A}\left( \theta ,\phi  \right)={{\mathbf{a}}_{\text{Rx}}}\left( \theta ,\phi  \right)\mathbf{a}_{\text{Tx}}^{H}\left( \theta ,\phi  \right),
\end{equation}
where ${{\mathbf{a}}_{\text{Tx}}}$ and ${{\mathbf{a}}_{\text{Rx}}}$ are the steering vectors at transmit array and receive array, associated with the angle-of-arrival (AoA) for reception and the angle-of-departure (AoD) for transmission. 

Generally, it is assumed that $P-1$ undesired targets are distributed throughout the environment, causing clutters. Thus, the targets can be indexed by $p \in \mathcal{P} \triangleq \{ 0,1,\ldots, P\}$, where $p=0$ represents the desired sensing target. 

Therefore, the channel impulse responses (CIRs) of sensing channel can be expressed as \cite{10334037,10078840 }
\begin{equation}
{{\mathbf{H}}^{\text{sen}}}\left( t,\tau  \right)=\sum\limits_{p=0}^{P}
{\begin{aligned}
		& \sqrt{{{G}_{p}}}{{e}^{j2\pi {{f}_{{{D}_{p}}}}\left( t \right)t}}{{e}^{j2\pi {{f}_{c}}{{\tau }_{p}}}}\cdot  \\ 
		& {{\mathbf{A}}_{\text{sen}}}\left( {{\theta }_{p}},{{\phi }_{p}} \right) \delta \left( \tau -{{\tau }_{p}} \right), \\ 
\end{aligned}}
\end{equation}
where ${{G}_{p}}=\frac{{{\lambda }^{2}}{{\sigma }_{RCS,p}}}{\left( 64{{\pi }^{3}}D_{p}^{4} \right)}$ is the fading coefficient related to the $p$th target, ${{f}_{{{D}_{p}}}}\left( t \right)= \frac{2{{v}_{p}}\left( t \right)}{\lambda }$ and ${{\tau }_{p}}=\frac{2{{D}_{p}}}{c}$ are the Doppler frequency shift and delay of the $p$th target, respectively, ${{\sigma }_{RCS,p}}$, $D_p$, and $v_p$ are the RCS, distance, and radial velocity of $p$th target, respectively. 	${{\mathbf{A}}_{\text{sen}}}\left( {{\theta }_{p}},{{\phi }_{p}} \right)$ denotes the product of the steering vector of transmit array and receive array of ISAC BS in target and environmental cluster sensing.
The key parameters and their definitions are given in Table \ref{sensing}.
\begin{table}[t]
	\setlength{\extrarowheight}{3pt}
	\centering
	\caption {Sensing channel modeling parameters}
	\label{sensing}
	\begin{tabular}{|l|p{6.3cm}|} 
		\hline
		\textbf{Parameters}     &    \textbf{Definitions}  \\ \hline	
				
		$G_p$          & The large-scale pass loss between BS and the target-$p$. \\ \hline
		$D_p$          & The distance between BS and the target-$p$.  \\\hline
		$f_{D_p}$      & The Doppler frequency shift of echo signal reflected from the target-$p$. \\\hline
		$\tau_p$       &  The delay of signal reflected from the target-$p$.  \\\hline
		$\left( \theta_p, \phi_p \right) $  &  Azimuth and elevation angles for the target-$p$. \\\hline
		${{\mathbf{A}}_{\text{sen}}}\left( {{\theta }_{p}},{{\phi }_{p}} \right)$       &  The steering vector used as target-$p$ sensing. \\\hline     
		
		\end{tabular}
\end{table}

\subsubsection{Communication Channel Modeling}
Without loss of generality, we assume that there are $L$ paths indexed by $ l\in \mathcal{L} \triangleq \left\{ 0, 1, \ldots , L \right\}$, including one LoS path $(l=0)$ and $L-1$ NLoS paths due to refraction, scattering, etc. Thus, the communication channel modeling of the $k$th user can be denoted by \cite{10334037,10078840}. 
\begin{align}
&\mathbf{H}_{k}^{\text{com}}\left( t,\tau  \right)=  \notag \\ 
& \begin{aligned}
		& \sqrt{\frac{K}{K+1}} G_0 {{e}^{j2\pi {{f}_{{{D}_{0}}}}\left( t \right)t}}{{e}^{j2\pi {{f}_{c}}{{\tau }_{0}}}}{{\mathbf{A}}_{\text{com}}}\left( {{\theta }_{0}},{{\phi }_{0}} \right) \delta \left( \tau -{{\tau }_{0}} \right)+ \\ 
		& \sum\limits_{l=1}^{L}{ \sqrt{\frac{1}{K+1}} G_l {{e}^{j2\pi {{f}_{{{D}_{l}}}}\left( t \right)t}} {{e}^{j2\pi {{f}_{c}}{{\tau }_{l}}}}\cdot {{\mathbf{A}}_{\text{com}}}\left( {{\theta }_{l}},{{\phi }_{l}} \right) \delta \left( \tau -{{\tau }_{l}} \right)}, \notag \\ 
	\end{aligned} \\ 
\end{align}
where $G_l$ is the large-scale fading, $K$ is the Rician factor, ${{f}_{{{D}_{0}}}}\left( t \right)=\frac{{{v}_{0}}\left( t \right)}{\lambda }$ and ${{\tau }_{0}}=\frac{{{D}_{0}}}{c}$ are the Doppler frequency shift and delay of the LoS path, respectively;  ${{f}_{{{D}_{l}}}}\left( t \right)=\frac{{{v}_{l}}\left( t \right)}{\lambda }$  and ${{\tau }_{l}}=\frac{{{D}_{l}}}{c}$ are the Doppler frequency shift and delay of the $l$th NLoS path, respectively; $D_0$, $D_l$, $v_o$, $v_l$ are the distances and radial velocities of the LoS path and the $l$th NLoS path, respectively. ${{\mathbf{A}}_{\text{com}}}$ denotes the product of the steering vector of transmit array of ISAC BS and receive array of user-$k$ through LoS and NLoS paths.
The key parameters and their definitions are given in Table \ref{communication}.
\begin{table}[t]
	\setlength{\extrarowheight}{3pt}
	\centering
	\caption {Communication channel modeling parameters}
	\label{communication}
	\begin{tabular}{|p{2cm}|p{6cm}|} 
		\hline
		\textbf{Parameters}     &    \textbf{Definitions}  \\ \hline	
		$G_l$          & The large-scale pass loss of $l$th path between BS and the user-$k$. \\ \hline
		$K$            & Rician factor. \\ \hline
		$f_{D_0} (f_{D_l})$      & The Doppler frequency shift of the LoS path ($l$th NLoS path). \\ \hline
		$\tau_0(\tau_l)$       & The delay of LoS path ($l$th NLoS path).  \\\hline
		$\left( \theta_0, \phi_0 \right), \left( \theta_l, \phi_l \right) $  &  Azimuth and elevation angles for UE ($l$th NLoS path). \\\hline
		${{\mathbf{A}}_{\text{com}}}\left( {{\theta }_{l}},{{\phi }_{l}} \right)$       &  The steering vector of BS and user-$k$ throhgh $l$th pass. \\\hline 
		
	\end{tabular}
\end{table}

\subsection{Performance Metric}
Communication performance in ISAC systems can be measured in terms of single-user rate, outage probability, and communication mutual information to evaluate the effectiveness and reliability of communication signal transmission. 
Whereas sensing performance can be measured using detection probability, resolution, accuracy, ambiguity function and sensing mutual information for the performance of detection and parameter estimation. 
Table \ref{metrics} summarizes the common performance metrics, combined with their definitions and descriptions for dual S\&C functionalities in ISAC systems. 

\setlength{\extrarowheight}{4pt}
\begin{table*}[htbp]
	\centering
	\caption { Important Performance Metrics of ISAC systems}
	\label{metrics}
	\begin{tabular}{|l|p{2cm}|p{0.6cm}|p{3.8cm}|p{4.4cm}|p{4cm}|}
		\hline
		\textbf{System}&\textbf{Metric} & \textbf{Paper} & \textbf{Definition} &\textbf{Meaning of Parameters} & \textbf{Description } \\ \hline			
		\multirow {3} {*} {Comm.}  
		&  Single-User Rate &  \cite{ 9739078}     & $ R=B\text{log}_2(1+SNR)$    & $ B$ is the signal bandwidth, and SNR is the power of received communication signal, typically expressed in decibels (dB). & This metric describes the maximum information rate. \\ \cline{2-6} 
		& Outage Probability  & \cite{ 7147772 } &  $P_d = P[R<R_{th}]$    &  $R$ represents the data transmission rate, and $R_{th}$ is the minimum data transmission rate threshold. & The probability that a communication system cannot achieve a specific data transmission rate  under given channel conditions.  \\ \cline{2-6}
		& Communication Mutual Information & \cite{  lin2018mutual} & $I(X; Y|H)$
		&  $H$ is the communication channel response, $Y$ is the received signal, and $X$ is the  transmitted signal.
		  &  This metric assesses the level of information sharing during the process of communication channel transmission.\\ \hline
		
		\multirow {5} {*} {Sens.}  
		& Detection Probability &  \cite{ma2008soft }     &  $P_D = P (\Gamma > \mu |H_1)$   & $\Gamma $ is the likelihood ratio under the hypothesis between a target is present $(H_1)$ and no target, and $\mu$ is the detection threshold based on the desired false alarm probability. & The probability of correctly identifying or detecting the target under given conditions.  \\ \cline{2-6} 
		& Resolution  & \cite{  book } &  Range resolution: $\Delta R= \frac{c}{2B}$. Velocity resolution: $\Delta v= \frac{\lambda}{2NT_r}$. Angular resolution: $\Delta \theta \approx \theta_{3 \text{dB}} = 0.886 \frac {\lambda}{D} $.    & $c$ is the speed of light, $ B$ is the signal bandwidth, $\lambda $ is the wavelength, $N$ is the pulse number, $T_r $ is the pulse repetition interval, $D$ is the aperture size of the antenna array, and $\theta_{3\text{dB}}$ is the half-power/$3$ dB beamwidth.  & The ability of a system to distinguish between two closely related targets. \\ \cline{2-6}
		& Accuracy &  \cite{9512486, xi2020joint }  & Range accuracy: $\sigma_R = \frac{\Delta R}{\sqrt{2SNR}}$. Velocity accuracy: $\sigma_v = \frac{\lambda}{2}\Delta f_d$. Angular accuracy: $\sigma_{\theta} = \frac{\theta_{3 \text{dB}}}{1.6\sqrt{2SNR}} $.  & $\Delta R$ is the range resolution, $\Delta f = \frac{1}{NT_r} $ is the Doppler resolution, SNR is the signal-to-noise ratio of echo. & Accuracy refers to the closeness of the measurement to the actual value of a quantity.  \\ \cline{2-6}
		& Ambiguity Function  & \cite{1163621} & $\chi(\tau, f_d) = \int_{-\infty}^{\infty} u(t) \cdot u^*(t - \tau) \cdot e^{j2\pi f_{d}t} \, dt$  & $\tau$ denotes the time delay,  $f_d = \frac{2v}{\lambda}$ denotes the Doppler frequency shift, and $u(t)$ represents the complex envelope of the transmitted signal. & Under specific signal processing conditions, the system's ability to resolve the distance and velocity of a target. \\ \cline{2-6}
		& Sensing Mutual Information & \cite{ ouyang2023integrated } & $I(G; Y|X)$  &  $G$ is the target response, $Y$ is the reflected echo, and $X$ is the sensing signal. &   The capability of extracting environmental information from the received signals. \\ \hline
		
	\end{tabular}
\end{table*}

\section{Single BS SI Management }

\begin{figure}
	\centering
	\includegraphics[width=0.9\linewidth]{ 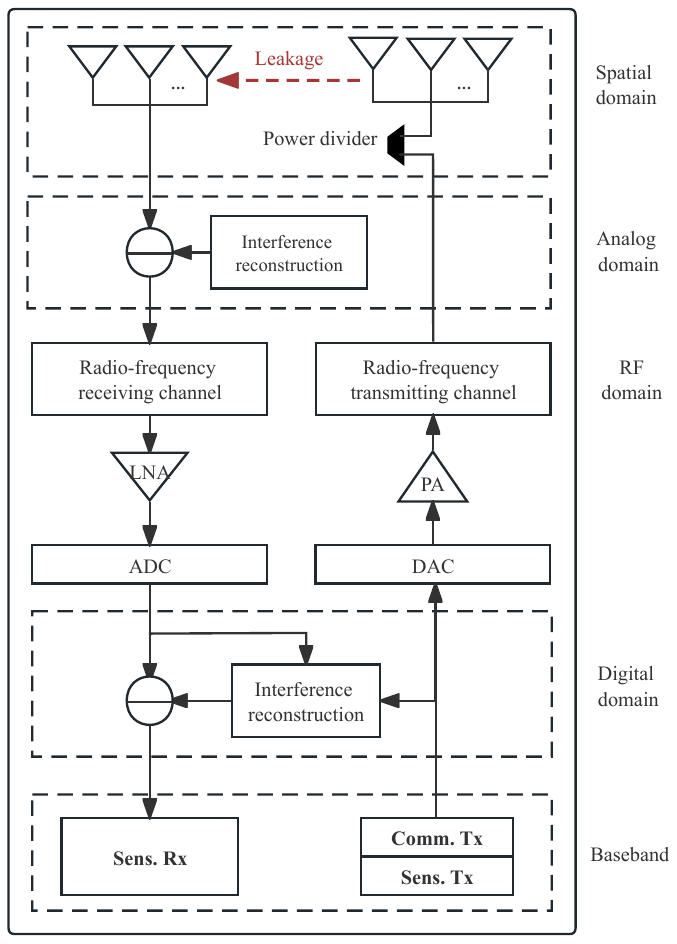}
	\caption{ The transmission process of baseband signal and the generation of SI \cite{9838784}. } 
	\label{SI}
\end{figure}
SI in wireless communication system refers to the signal leaking from transmitter to the receiver, as illustrated in Fig. \ref{SI}. The SI can be categorized into analog SI and digital SI, which occurs in the analog domain and digital domain, respectively. 
The foundation of analog SI lies in the frequency-selective transmit-receive leakage paths within the analog front end. 
For instance, leakage occurs when the power amplifiers (PA) at the transmitter, the low noise amplifier (LNA), and the mixer at the receiver operate in non-linear regions.
Digital SI arises from frequency aliasing during the baseband signal sampling process and system clock synchronization, primarily affecting the digital signal processing.

Generally, the SI is so large that it saturates the analog-to-digital converter (ADC) in the receiver, which leads to \textit{(i)} non-linear distortion and \textit{(ii)} harmonic and intermodulation distortion, causing the loss of communication data. 
Moreover, saturation constrains the dynamic range of ADC, making it challenging to process weak signals and sense the distant targets.

Therefore, it is critical to suppress SI and the objective is to reduce it to the noise floor of the local receiving channel. Due to the signal processing in different domains, SI suppression is categorized into analog SI suppression and digital SI suppression \cite{9838784}. 
Moreover, SI can be suppressed in the spatial domain before signal processing in the receiving circuit. Thus, this section sequentially reviews SI suppression and avoidance methods in ISAC systems within a single BS. A summary of SI management is provided in Table \ref{Summary SI}. 

\setlength{\extrarowheight}{4pt}
\begin{table*}[htbp]
	\centering
	\caption { Summary of SI Management Methods}
	\label{Summary SI}
	\begin{tabular}{|l|l|l|p{9cm}|}
		\hline
\textbf{Scopes}&\textbf{Specific Methods}  
& \textbf{References} & \textbf{Summary in one sentence} \\ \hline			
\multirow {2} {*} {Analog Domain} 
&  Hardware Parameter Optimization  &  \cite{morgenstern2022analog, morgenstern2023analog, 7850994}   &  Reconstruct and eliminate SI based on the known structure and transmission characteristics of the interference signal.  \\ \cline{2-4}
& Analog Cancellation Algorithm   & \cite{8053906} & A two-stage SI cancellation algorithm combines the dual-tap analog cancellation circuit and nonlinear model. \\ \hline

\multirow {5} {*} {Digital Domain}  
&  Add Auxiliary Links  &  \cite{8805161, 7051286, 7941987}     & Auxiliary receiver chain can reduce the impact of phase noise at the receiver. \\ \cline{2-4}
& Transmit Antenna Design   & \cite{9339925, 8169087,9363029} &  Eliminate SI by flexibly selecting antennas and designing transceiver beamforming matrices. \\ \cline{2-4}
& Transmitted Signal Design  & \cite{xiao2021full, 9565357} & Design signal transmission format based on sequence, pulse, orthogonal frequency division multiplexing (OFDM), etc. \\ \cline{2-4}
& Digital Cancellation Algorithm  & \cite{9395095, 8249154, 7760645} &Reconstruct and eliminate digital SI signals at the receiver's RF front-end before radar signal processing. \\ \cline{2-4}
& Receive Filter Design & \cite{8449215, 8259307} & Design matched filters and sample the echo signals when SI is at its weakest. \\ \hline

\multirow {3} {*} {Broadcast Domain}  
&  Hardware Design        &  \cite{4982861, WOS:000649996200001, 10.5555/3066994}   &  Physical isolation is employed to prevent leakage of transmitted signals.    \\ \cline{2-4}
& Antenna Structure Design   & \cite{9597643} & A two-stage SI cancellation algorithm,  comprises a dual-tap analog cancellation circuit and nonlinear model. \\ \cline{2-4}
& Beamforming Optimization   & \cite{7549068,8979472, 5985554} & Isolation-based antenna structure design can significantly reduce the coupling phenomena at the transmitter and receiver.\\ \hline
		
	\end{tabular}
\end{table*}

\subsection{SI Suppression}
\subsubsection{Analog SI Mitigation}
Utilizing ``control-constrained hardware'' refers to hardware with time requirements, which is necessary for analog interference mitigation,  because the resolution of radio frequency (RF) components critically affects the effectiveness of analog SI suppression \cite{kwak2022analog}.
Additionally, accurate sensing requires ISAC signal of cyclostationary characteristics, resulting in residual SI inevitably. 
Therefore, this subsection reviews the methods for suppressing analog SI from hardware parameter optimization and analog cancellation algorithm.
	\begin{itemize}
	\item \textbf{Hardware Parameter Optimization: } Hardware parameter optimization refers to the method of optimizing the hardware design to achieve SI suppression based on the characteristics of the interfering signal. 
	Morgenstern \textit {et al.} minimized the residual SI by using the LNA's output signal as a reference and optimizing tap weights \cite{morgenstern2022analog}. 
	Under hardware constraints, this method can achieve over $30$ dB of SI suppression at lower frequency bandwidths. However, the research mentioned above only focuses on tap spacing and does not analyze the relationship between hardware design and hybrid tap filter. 
	Furthermore, Morgenstern \textit {et al.} introduced a method that characterizes hardware constraints such as channel bandwidth, carrier frequency, and bit precision, which enhances the tap-delay resolution and achieves up to $45$ dB of SI suppression in analog domain \cite{morgenstern2023analog}. 
	Additionally, Huang \textit{et al.} proposed a multi-tap adaptive filter by regarding an analog least mean-square (ALMS) loop as a part of the RF front-end to enhance the SI suppression, because the auto-correlation of transmitted signal and loop gain are vital factors in interference channel modeling \cite{7850994}. 
			
	\item \textbf{Analog Cancellation Algorithm:} 
	SI cancellation in the analog domain allows the SI signal to be directly reconstructed in the RF domain and subtracted from the received signal, which avoids saturation of the ADC at the receiver and reduces the quantization noise. However, analog circuits are susceptible to non-linear factors, such as the non-linear effects of PAs, which can lead to higher-order interference. Moreover, the design and optimization of analog circuits require complex hardware tuning.
	 Liu \textit{et al.} proposed a full-duplex transceiver architecture employing a two-stage analog SI cancellation structure \cite{8053906}. 
	Specifically, the first stage utilizes a dual-tap analog cancellation circuit to alleviate SI caused by direct leakage and a reflected component, thereby mitigating the impact of multipath SI on the dynamic range. The second stage employs a nonlinear model to compensate for the residual SI left in the first stage, thereby enhancing the overall performance of analog cancellation. 
		
 	\end{itemize} 
	
\subsubsection{Digital Interference Mitigation}
Digital SI cancellation, which operates on signals after ADC conversion, is more flexible and less constrained by hardware limitations. SI cancellation in digital domain can be achieved by adding auxiliary links, transmit antenna design, transmit signal design, digital cancellation algorithms and receive filter design.

\begin{itemize}
	\item \textbf{Add Auxiliary Links:} 
	Digital SI cancellation technology, while offering low complexity for SI mitigation, faces limitations in interference resistance due to the nonlinearity of local transmitter components in traditional approaches \cite{8805161}. 
	In \cite{7051286}, an auxiliary receiver chain obtains a digital-domain copy of the transmitted RF SI signal. By sharing an oscillator with the primary receiver, the auxiliary receiver reduces the impact of phase noise on the receiver, thereby enhancing the gain of SI cancellation. 
	Li \textit{et al.} incorporated an auxiliary chain and exploited the independence between the echo signals and SI signals to develop two interference cancellation algorithms based on independent component analysis. These algorithms extract the desired signal directly from the received signal, significantly reducing computational complexity compared to the traditional least-squares (LS) approach \cite{7941987}. 
	
	\item \textbf{Transmit Antenna Design:}
	Effective suppression of SI in the digital domain can be achieved by flexibly selecting antenna and transmit beamforming designs. However, this method is sensitive to environmental conditions and requires frequent calibration and adjustment.
	Besides, the method is sensitive to the antenna array and environmental conditions, thus requiring frequent calibration and adjustment.
	 Huang \textit{et al.} proposed a method to eliminate SI by flexibly selecting antennas and designing transmit-receive beamforming, leveraging the known characteristic distribution of the SI channel. 
	Specifically, the authors analyzed the explicit relationship between the interference suppression ratio (ISR) and channel estimation errors, then the performance of successive interference cancellation (SIC) can be directly evaluated through ISR \cite{9339925}. 
	From a signal processing perspective, exploring new beamforming schemes is also a popular design approach. For instance, B. Nuss \textit{et al.} combined digital beamforming technology on an integrated BS to focus interference in the angular domain towards the arrival direction of the interference source, enabling interference cancellation in other directions \cite{8169087, 9363029}.

	\item  \textbf{Transmitted Signal Design:} 
Pulse-based ISAC signal design operates by transmitting and receiving signals at different times and can achieve complete SI cancellation. Typically, the longer the repetition cycle of the transmitted signal, the less likely SI leakage occurs. However, this approach results in underutilization of spectrum.
To address this contradiction, Xiao \textit {et al.} proposed a full-duplex ISAC system that incorporates SI leakage, where the waiting time of conventional pulse radar is used to transmit dedicated communication signals. This approach not only suppresses SI but also utilizes the dedicated communication signals to enhance sensing performance \cite{xiao2021full}. 
Therefore, communication signals can also be employed for target sensing. For instance, Tang \textit {et al.} introduced an SI calibration algorithm that leverages the consistent Golay sequence in the preamble of communication signals. This algorithm can subtract the SI signal in the digital domain to combat SI in short-range sensing. Moreover, the authors  proposed a signal design scheme for distant targets by evenly inserting pulse radar or reference signal subcarriers on OFDM subcarriers, which prevents residual SI affecting the signal-to-noise ratio (SNR) \cite{9565357}.
	
	\item \textbf{Digital Cancellation Algorithms:} 
The Digital SI cancellation algorithm is achieved through a two-step process involving reconstructing and eliminating digital domain SI signals from the receiver's RF front-end before radar signal processing. Therefore, the acquisition of the interference signal, particularly its nonlinear characteristics, plays a crucial role in the effectiveness of interference cancellation.
To address the issue of nonlinear SI, Wang \textit{et al.} proposed a two-part nonlinear SI cancellation scheme that first uses a neural network to capture the nonlinear characteristics, followed by a linear filter to capture the linear characteristics. This approach achieves over $20$ dB of interference cancellation \cite{9395095}. 
In addition, to mitigate the loss of radar signals during interference separation, interference cancellation techniques often incorporate sparse signal recovery algorithms to minimize the deficit in the SNR of the sensing process \cite{8249154}.
For instance, Hakobyan \textit{et al.} employed a linear prediction approach to recover excluded subcarrier signal values by utilizing forward and backward linear predictions from values of adjacent subcarriers. The signal-to-interference ratio (SIR) can be improved by $12-16$ dB \cite{7760645}.   

\item \textbf{Receive Filter Design:} 
Matched filters can effectively reduce the peak response of the ALMS loop, thereby significantly mitigating the frequency component of residual SI in the digital domain. Le \textit{et al.} further noted that the performance of the ALMS loop in the digital domain is determined by four elements: loop gain, tap delay, number of taps, and roll-off factor, which provides a theoretical basis for designers to determine the cancellation level required for the ALMS loop \cite{8449215}.
In \cite{8259307}, a receiver structure is proposed to address the significant power asymmetry between SI and the desired signal by sampling the echo signals when the interference is at zero amplitude. While this approach preserves the desired signal, it requires knowledge of the interference signal, which is incredibly challenging.
	
\end{itemize}

\subsection{ SI Avoidance}

In addition to eliminating SI signals in the analog and digital domains, providing sufficient isolation in the propagation domain for the transceiver can indirectly reduce subsequent computational complexity. Therefore, this section will review SI avoidance methods from the perspectives of hardware design, antenna structure design, and beamforming optimization.
\begin{itemize}
	\item \textbf{Hardware Design: } 
Hardware structures, including inductors, circulators, ortho-couplers, and acoustic filters, can be employed to suppress the generation of SI. In \cite{4982861}, eight-shape inductors are used to reduce the coupling. Below the inductors, shields can be inserted to improve their quality factors and prevent the substrate from coupling. Therefore, this method can enhance the linearity of power amplifier drivers (PADs), thereby avoiding the generation of SI signals. However, the improvement in linearity often comes with increased power consumption, which is typically undesirable in cellular systems \cite{WOS:000649996200001}.
To this end,  Krikidis \textit{et al.} employed two lumped-element circulators and a $3$ dB quadrature coupler to suppress the generation of SI signals, resulting in enhanced communication transmission rates without increasing power consumption \cite{10.5555/3066994}. 

\item \textbf{Antenna Structure Design: } 
Isolation-based antenna structure design can significantly reduce coupling phenomena at the transmitter and receiver. In \cite{9597643}, the authors introduced an antenna cancellation technique employing a symmetric antenna structure and a $\pi$ phase shifter to mitigate destructive interference for enhanced isolation. 

\item \textbf{Beamforming Optimization: } 
With the transmit beamforming SIC technique, signals from the transmit antennas are transmitted into the null space of the SI channels so that they are nulled when arriving at the receive antennas \cite{7549068,8979472}. 
However, in practical applications, factors such as noise, nonlinearity, implementation of estimation algorithms, and hardware limitations can introduce channel estimation errors and other defects that affect the performance of SIC. To address this issue, Riihonen \textit{et al.} conducted simulations to investigate the impact of channel estimation errors on SI suppression in transmit beamforming \cite{5985554}. 
As system complexity grows, the fundamental approach to beamforming design remains consistent. However, there is a notable increase in computational complexity. 
	
\end{itemize}

\section{ Single BS MI Management}
Simultaneously accomplishing S\&C tasks presents unique challenges in signal transmission and sensing processing.
Due to spectrum resource sharing and mutual assistance between S\&C functions, sensing signals and communication signals entirely or partially overlap in the time and/or frequency domains, resulting in MI. 
MI diminishes both network throughput in communication and dynamic range in sensing, thereby impacting the overall performance of ISAC systems.
The generation of MI involves various stages, such as signal design, transmission, and reception. It is also an inherent challenge in achieving performance trade-offs between S\&C dual functions. 
In this section, a review of MI suppression, avoidance, and exploitation in the uplink and downlink stages will be conducted separately.

\subsection{Downlink MI Suppression}
\begin{figure}
	\centering
	\includegraphics[width=0.8\linewidth]{ 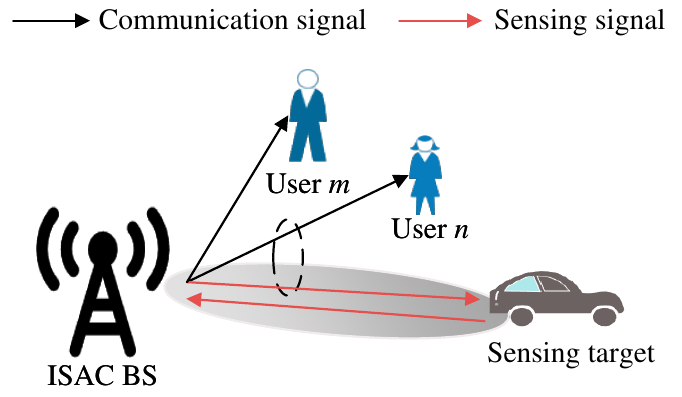}
	\caption{ The illustration of downlink MI. } 
	\label{downlink}
\end{figure}
As depicted in Fig. \ref{downlink}, sensing signals and downlink communication signals  interfere with each other.
Downlink MI suppression consists of directly reconstructing and subtracting the interference signals, and indirectly altering the direction of interference signals to distant subspaces from the desired signal.
Table \ref{downlink MI} provides a concise summary. 

\setlength{\extrarowheight}{4pt}
\begin{table*}[htbp]
	\centering
	\caption { Summary of Downlink MI Management Methods}
	\label{downlink MI}
	\begin{tabular}{|l|l|p{1.5cm}|p{9cm}|}
		\hline
		\textbf{Scopes}&\textbf{Specific Methods}  
		& \textbf{References} & \textbf{Summary in one sentence} \\ \hline			
		\multirow {2} {*} {MI suppression} 
		& Direct Interference Suppression  &  \cite{8025554, 5960632}   &  Physical isolation is employed to prevent leakage of transmitted signals.    \\ \cline{2-4}
		& Indirect Interference Suppression & \cite{9024542, 8239809}   & Detect, classify, and pinpoint the source of interference to subsequently mitigate MI. \\ \hline
		
		\multirow {4} {*} {MI Avoidance}  
		&  Multiple Access Techniques &  \cite{8917703, 10283528, 9728752, zhang2023semi, 9771628, 10318068, 9656537, 9557830, 10024901, 7414384, 9761984, 9831440, li2024finite, 9531484}     & Suppress MI and enhance resource utilization efficiency in the time, frequency, spatial, and resource domains by designing MA techniques. \\ \cline{2-4} 
		& Beamforming Design  & \cite{9124713, 8386661, 9724206,9534484} & Weighted combinations of separate radar waveforms and communication symbols are precoded to create multiple beams directed towards radar targets and communication receivers. \\ \cline{2-4}
		& Transmitted Signal Design  & \cite{9839026,10192417,8043792,9668964,10109100,9036061,6104176,9124713,8386661} & Design transmission signals according to requirements, including sensing-centric design, communication-centric design, and weighted design. \\ \cline{2-4}
		& RIS Assisted & \cite{10042240, 10052711, 10184278} & Customize the environment to facilitate signal transmission and avoid MI. \\ \hline
		
		\multirow {2} {*} {MI Exploitation} 
		& Power Exploitation  &  \cite{7103338,9593096,9465648,9534484,9585492}   &  Exploit MI to enhance received signal power or reduce transmitted signal power.   \\ \cline{2-4}
		& Reduce Symbol Error Rate & \cite{6619580,8355705, 7103338, 10005221}   & Shift the received interference-free signal away from the decision threshold. \\ \hline
		
	\end{tabular}
\end{table*}

\subsubsection{Direct Interference Suppression}
Direct interference suppression relies on the known structure and transmission characteristics of interference signals, enabling the reconstruction and subsequent subtraction of MI signals. 
For example, by combining prior information to model interference channels, signals with amplitudes exceeding the strongest interference threshold can be filtered out and subtracted \cite{8025554}. 
Decoders based on pilot signal techniques are designed to detect sensing echoes and eliminate interference. Additionally, useful communication signals can be extracted from regularly spaced pilot symbols, facilitating interference signal reconstruction using pilot signals to enhance radar dynamic range \cite{5960632}. 

\subsubsection{ Indirect Interference Suppression}
Indirect interference cancellation alters or anticipates the transmission direction of interference signals in the next time slot, thereby causing the interference signals to propagate within subspaces distant from the desired signal. 
For instance, when prior channel information is unknown, Wu \textit{et al.} utilized a recursive neural network-based autoencoder for blind interference detection \cite{9024542}. This method reconstructs the transmission signal and automatically detects, identifies, and locates the source of interference, thereby anticipating the propagation subspace of MI.
Moreover, Jing \textit{et al.} integrated spatial and temporal dimensions and introduced a linear space-time interference alignment algorithm tailored for multi-user MIMO interference channels \cite{8239809}. 
This approach effectively reduces the number of antennas required for complete interference elimination, demonstrating robustness in communication scenarios with multiple interference sources and high interference levels. 
However, the effectiveness of these methods in interference suppression depends on known information, such as interference channel modeling and pilot structures, leading to limitations in flexibility. 

\subsection{Downlink MI Avoidance}

\subsubsection{Multiple Access (MA) Techniques} 
\begin{figure}
	\centering
	\includegraphics[width=0.8\linewidth]{ 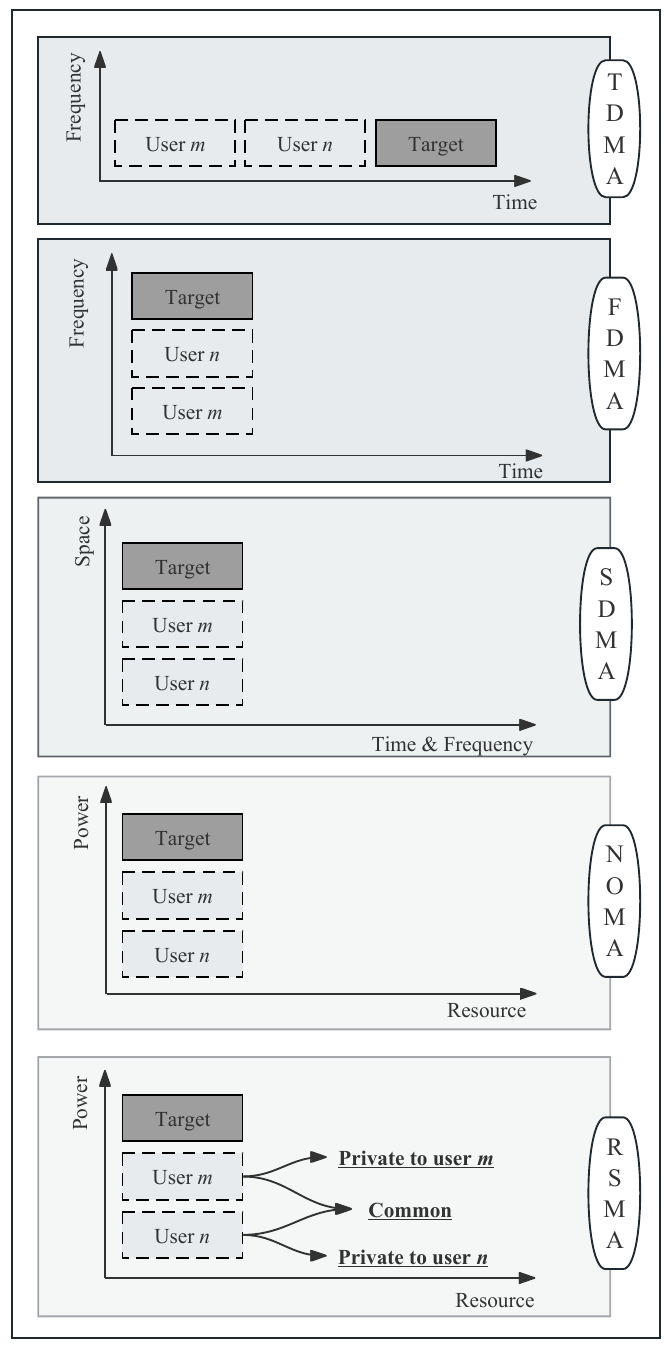}
	\caption{ The resource allocation diagrams in MA techniques. } 
	\label{MA}
\end{figure}
MA is a feasible solution in ISAC systems to suppress interference and enhance resource utilization efficiency.
As illustrated in Fig. \ref{MA}, a comparison of various MA techniques is presented, including time division multiple access (TDMA), frequency division multiple access (FDMA), non-orthogonal multiple access (NOMA), spatial division multiple access (SDMA), and rate-splitting multiple access (RSMA). 
The advantages of these MA techniques in avoiding MI are summarized as follows.

\begin{itemize}
\item \textbf{TDMA:} 
Frequent switching between pilot signals and infrastructure is necessary in time division systems to obtain accurate channel state information (CSI) for establishing and maintaining communication links.
Consequently, communication throughput is severely restricted by the association overhead. 
To tackle this challenge,  Kumari \textit{et al.} proposed a dynamic slot allocation method between S\&C that achieves performance trade-offs \cite{8917703}.
In addition, Li \textit {et al.} designed a novel frame structure that reallocates the traditional guard period and uplink feedback period resources in new radio (NR) wireless communications to enhance downlink communication and sensing functions.
Under the specified NR frame structure, communication overhead can be reduced by $43.24$ \% \cite{10283528}. 
    To address the decrease in target localization accuracy caused by the overhead of sharing sensing data, Zhang \textit{et al.} proposed a flexible time-frequency resource allocation scheme and evaluated the accuracy of sensing data fusion through analysis of response delay. The actual hardware platform test data demonstrates that, in the mmWave frequency band, this approach reduces the target detection error by $18.5$ \% \cite{9728752}.
	
	\item \textbf{FDMA:} 
	The entire frequency band occupied by the ISAC system can be divided into three parts: dedicated communication signals, dedicated radar signals, and ISAC signals \cite{zhang2023semi}. 
	Therefore, allocating channels based on packet arrival rates rather than assigning fixed frequency bands to specific functions can significantly reduce MI and enhance spectrum resource utilization efficiency \cite{9771628}. 
	Compared with uniform power allocation, a lower Cram\'{e}r-Rao lower bound (CRLB) can be achieved by sparsely selecting the optimal sensing subcarriers and allocating the optimal transmit power.
	Wu \textit{et al.} highlighted that the MI between S\&C functions can be effectively mitigated by establishing synchronization links at distinct frequency bands \cite{9656537}. 
	
	\item \textbf{NOMA:} 
	NOMA enables multiple UEs to be efficiently served on the same wireless resources, thereby improving resource utilization efficiency. Therefore, applying NOMA to ISAC offers unique advantages in managing interference between users and inter-functionality interference \cite{9557830}. 
	Even with limited SNR diversity gain, the reliability of the system under the ideal condition is significantly improved compared to the imperfect successive interference cancellation mode \cite{li2024finite}. 
	Specifically, Mu \textit{et al.} proposed two novel designs to effectively coordinate interference among users and sensing against communication interference, namely NOMA-empowered downlink ISAC and NOMA-inspired downlink ISAC \cite{10024901}.  
	\begin{figure}
		\centering
		\includegraphics[width=1\linewidth]{ 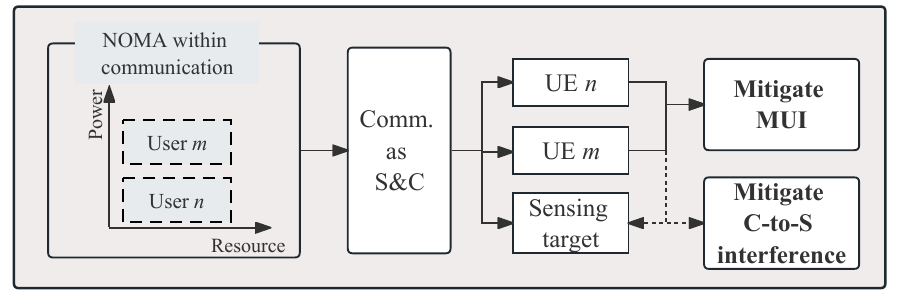}
		\caption{ The NOMA-empowered downlink ISAC design \cite{10024901}. } 
		\label{NOMA1}
	\end{figure}
	\begin{figure}
	\centering
	\includegraphics[width=1\linewidth]{ 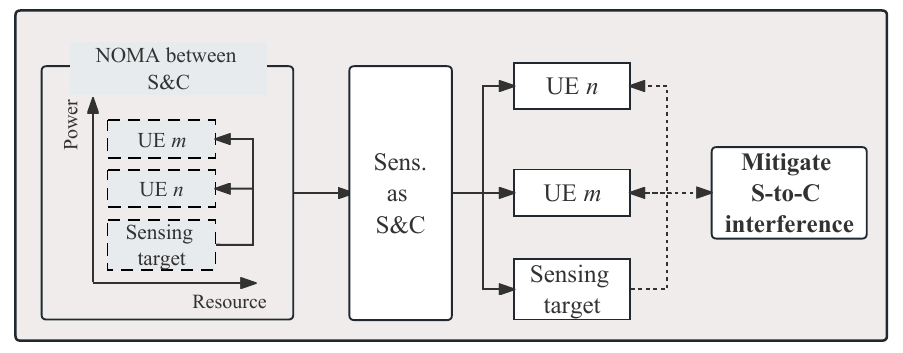}
	\caption{ The NOMA-inspired downlink ISAC design \cite{10024901}. } 
	\label{NOMA2}
\end{figure}
	\paragraph{NOMA-empowered ISAC} 
In NOMA-empowered downlink ISAC, NOMA is applied among UEs, with superposition coding (SC) and SIC techniques for transmitting and detecting the communication signals at each user, which enables the same superimposed communication signals for sensing, preventing additional interference to communication. 
This concept is illustrated in Fig. \ref{NOMA1}. 
	
	\paragraph{NOMA-inspired ISAC} 
	In NOMA-inspired downlink ISAC, additional sensing signals are utilized for communication to form extra joint S\&C waveforms.
	At each UE, these additional waveforms are initially detected and eliminated through SIC, thereby mitigating the interference from sensing to communication in the downlink ISAC, as shown in Fig. \ref{NOMA2}.
	
	\item \textbf{SDMA:} 
    The extensive use of multi-antenna technology in ISAC systems enhances signal strength and reduces interference. In SDMA systems, multiple UEs and targets can be served or sensed in the same time/frequency/code domain while reducing MI through spatially distinct beams. Thus, beamforming design plays a critical role in interference management \cite{7414384}.
  	For example, He \textit{et al.} reduced the MI by optimizing the beamforming design, where the radar transmission can be oriented to the desired target direction while ensuring minimal interference to the communication signals \cite{9761984}.
	
	\item \textbf{RSMA:} 
	RSMA divides communication signals into common and private components. The common signals are collectively encoded into a shared stream for decoding by multiple users. In contrast, the private signals are independently encoded into individual streams for decoding by particular users. 
	Therefore, the rate splitting concept partially treats interference as noise and requires only partial interference decoding, thereby contributing to higher spectral efficiency \cite{9831440}.
	For example, Xu \textit{et al.} partitioned the information into common and private streams at the transmitter and utilized the common portion for sensing without additional radar sequences \cite{9531484}. 
	Since the sensing signal is a part of the communication signal, this approach not only enhances the radar beamforming but also reduces the interference of sensing on communication. 
	Furthermore, Gao \textit{et al.} utilized both the common and private signals for sensing through a cloud-radio access network (C-RAN). The authors characterized the achievable performance region in networks centered around communication and sensing, validating the advantages of RSMA for interference management and direct localization \cite{10032141}.
	
\end{itemize}

\subsubsection{	Beamforming Design} 
In this section, two beamforming designs are introduced: Firstly, distinct beams handle sensing and communication tasks, with sensing beams scanning periodically and communication beams maintaining high directionality. 
Secondly, main lobe and side lobe perform different functions and prevent ISAC system performance degradation by balancing the impact of side lobe on main lobe.
\begin{itemize}
	\item \textbf{ Different beams execute S\&C functions: } 
	The sensing beam periodically scans within a certain range to obtain target parameters, while the communication link is highly directional \cite{WOS:000457302900054}. 
	Therefore, downlink communication is susceptible to the influence of sensing. MI occurs when there is signal leakage between beams or partial overlap in coverage areas.
	Liu \textit{et al.} proposed a dual-function radar-communication (DFRC) system design that utilizes antenna array to transmit weighted combinations of independent radar waveforms and communication signals, illuminating the radar targets and communication users by spatial beams. Compared to previous methods that synthesized radar transmission beams by transmitting only communication symbols, this approach provides greater flexibility for MIMO radar, significantly improving radar performance \cite{9124713}.
	In \cite{8386661}, the authors introduced a beamforming optimization scheme for fully digital MIMO systems to minimize multi-user interference (MUI) energy and the mismatch error between the desired and transmitted beams.
 	
	\item \textbf{The main lobe and side lobe perform different functions: } 
	While suppressing side-lobe interference, it is vital to balance main-lobe functionality to prevent a sharp decline in the performance of ISAC systems when MI leaks into the main lobe. 
	In this context, Chen \textit{et al.} strived to optimize beamforming to concentrate beam transmission energy within the most effective spatial sector, thereby mitigating the impact of MI signals \cite{9724206}.
	Additionally, Liu \textit{et al.} introduced a symbol-level precoding technology into ISAC systems, allowing for the transmission of ideal waveforms in each time slot. This innovation enhances the flexibility in waveform design \cite{9534484}.
	
\end{itemize}

\subsubsection{Transmited Signal Design} 

Transmitted signal design can be categorized into three types, namely sensing-centric design, communication-centric design, and weighted design.
\begin{itemize}
	\item \textbf{Sensing-centric transmitted signal design: } 	
	The main challenge of embedding communication signals into pulse or continuous wave signals lies in minimizing interference with sensing \cite{7746569,8828016,8828023}.
	For instance, Wang \textit{et al.} proposed to split the radar signals into pure radar signals and a combination of radar and communication signals. Thus, the sensing signal is extracted and utilized as a virtual communication signal, allowing SIC to eliminate co-channel sensing interference at the communication user\cite{9839026}.
	Additionally, a new ISAC signaling approach was proposed in \cite{10192417}, where the radar symbols are alternately reversed in consecutive periods and the data symbols are phase-rotated. 
	Therefore, the receiver processes signals every 2$T$ instead of every $T$, effectively reducing co-channel sensing interference since the signal parameters remain unchanged between adjacent radar symbol periods \cite{8043792}.
	
	\item \textbf{Communication-centric transmitted signal design: }
	Communication-centric ISAC signal design involves leveraging communication signals directly for sensing purposes. 
	For instance, Wang \textit{et al.} proposed a NOMA-ISAC framework, in which the transmitted superimposed signal is exploited for communication and sensing simultaneously and SIC is
	exploited for inter-user interference mitigation \cite{9668964}. 
	Furthermore, as indicated in \cite {10109100}, the communication signal's preamble sequence exhibits favorable autocorrelation properties, thus facilitating the identification and localization of the target. 
	However, within the coherent processing interval (CPI), the quantity of preamble sequences positively correlates with sensing effectiveness. 
	Nonetheless, communication efficiency inversely relates to the preamble sequence length, due to occupied communication data bits. In this regard, Chu \textit{et al.} developed a task-driven signal design approach to optimize the length of the preamble sequence within the CPI, thereby improving the sensing accuracy \cite{10109100}. 
	
	\item \textbf{Weighted transmitted signal design: } 
	In weighted transmitted signal design, the finer constraints imply an increased ability to fulfill the complicated sensing requirements in ISAC applications. 
	However, constraints will introduce additional complexity and challenges to the hardware implementation. 
	Therefore, appropriate constraints is essential in ensuring the validity of relaxed solutions and achieving a better trad-off between S\&C functions \cite {9036061}.  
	The typical constraints include power constraint, constant modulus (CM) \cite{6194367}, peak-to-average power ratio (PAPR) \cite{6104176}, and similarity constraint (SC) \cite{WOS:000330291700007}.  
	Specifically, Wang \textit{et al.} presented an efficient optimization method to design a CM probing signal which can synthesize a desired beam pattern while maximally suppressing both the auto-correlation and cross-correlation sidelobes at/between given spacial angles \cite{6194367}. 
	Moreover, Liu \textit{et al.} proposed a joint transmit beamforming model for a dual-function MIMO-ISAC system under the CM and SC constraints, while guaranteeing the signal-to-interference-plus-noise ratio (SINR) at each communication user \cite{8386661}. A reduced complexity algorithm is then proposed based on zero-forcing the inter-user interference and radar interference. 
 
\end{itemize}

    \subsubsection{RIS Assisted} RIS can improve wireless communication capacity and coverage by creating favorable propagation environments. Thus, RIS-based ISAC system allows for dynamic control of MI avoidance. Depending on whether signal amplification functionality is included, RIS can be categorized into active RIS and passive RIS. 
    
	\begin{itemize}
		\item \textbf{Passive RIS:} 
		Passive RIS consists of numerous passive elements, each capable of reflecting incident signals with controllable phase shift. 
		Therefore, MI can be avoided and the desired signals can be coherently superimposed with the same phase to obtain a higher array gain. 
		For example, Zhong \textit{et al.} proposed a comprehensive waveform design based on passive RIS to jointly maximize sensing SINR and minimize interference among multiple users in which the trade-off between S\&C is adjusted by a weight parameter. Simulation demonstrates that flexible waveform design outperforms systems without RIS in terms of communication rate \cite{10042240}. 		
		Furthermore, Luo \textit{et al.} focused on a more typical scenario where a dual-functional ISAC BS continuously detects multiple targets and simultaneously serves multiple users through traditional links and RIS-assisted links. Under the requirements of communication quality of service (QoS), total transmission power budget, and RIS phase constraints, they jointly optimized the transmit beamforming and RIS reflection coefficients to maximize the target detection SINR \cite{10052711}.
	
	\item \textbf{Active RIS:} 
	Active RIS can compensate for the large path loss in RIS-assisted links by amplifying the reflected signals through amplifiers integrated into its components. 		
	In \cite{10184278}, a dual-functional ISAC BS is equipped with a uniform linear array (ULA) featuring delay alignment modulation (DAM) and an active RIS component composed of several reflective units. Particularly, DAM intentionally introduces delays at the BS to ensure that multipath signals arrive simultaneously at the receiver, thereby eliminating inter-symbol interference (ISI).  
	\end{itemize}
	
	\subsection{ Downlink MI Exploitation}
      
     In contrast to destructive interference, constructive interference (CI) refers to some of the MI signals that can help with enhancing the received SINR, and reducing the symbol error rate (SER) and transmitted signal power. In this regard, the exploitation of CI can
     improve the energy-efficiency (EE) of ISAC systems, which aligns with the concept of green communication \cite{7103338}. 
     The exploitation of the downlink MI signal with respect to power exploitation and SER reduction is provided in this subsection.
     
     \subsubsection{Power Exploitation}
    Power exploitation can be divided into enhancing the received SINR and reducing the desired transmit power.
\begin{itemize}
	\item \textbf{Enhance the received SINR: } 	
	Masouros \textit{et al.} proposed a precoding algorithm that selectively retains MUI beneficial to UEs, thereby enhancing the SNR at the receiver \cite{4286950}. 
	Su \textit{et al.} introduced an interference rotation technique that rotates the signals received by users into the decision region, thereby utilizing interference between multiple users to reduce the transmit power \cite{9593096}. 
	However, the above methods exploit the existing signal energy without actively introducing novel energy sources. 
	In contrast, Li \textit{et al.} proposed to use the IoT device to collect energy from a nearby power station via a direct link as well as a reflective link from the RIS to enhance the uplink between the IoT device and the access point (AP) \cite{9686954}. 
	This method achieves higher data transmission efficiency with limited energy harvesting, and also reduces the interference to APs, resulting in an increased throughput of the network.
 
	\item \textbf{Reduce the expected power of the transmitted signal: } 
	Under the same total transmit power constraint, Hong \textit{et al.} proposed a data-aided transmission precoding scheme, where beneficial MI can be obtained based on the known data and CSI sent by the transmitter, thereby reducing the expected power of transmitted signals \cite{9465648}. 
	Additionally, in \cite{9534484}, the conversion of harmful MUI signals into useful signals significantly improves the performance of simultaneous multi-user communication. 
	To address the uncertainties in user power constellation diagrams, Zhang \textit{et al.} proposed a waveform optimization scheme based on a Pareto optimization framework, which outperforms existing LS methods in terms of SER \cite{9585492}.
	
\end{itemize}

\subsubsection{Reduce SER}
In a multi-user MIMO system using phase-shift keying modulation, CI can assist UEs in the downlink with symbol decisions. 
Similarly, inter-channel interference and known MUI can be used to boost the received SINR for downlink users while minimizing disruption to sensing functions \cite{6619580,8355705}. 
Additionally, with known destructive interference, optimization techniques can rotate destructive interference into constructive areas, which improve the target detection probability and EE \cite{7103338}. 
In \cite{10005221}, a waveform optimization scheme is considered, where MI between sensing target and UEs are recognized as CI. By designing the transmit beamforming and sensing waveforms, the received noiseless signals are pushed as far away from the decision threshold as possible, significantly reducing the SER.

\subsection{Uplink MI Management}
\begin{figure}
	\centering
	\includegraphics[width=0.8\linewidth]{ 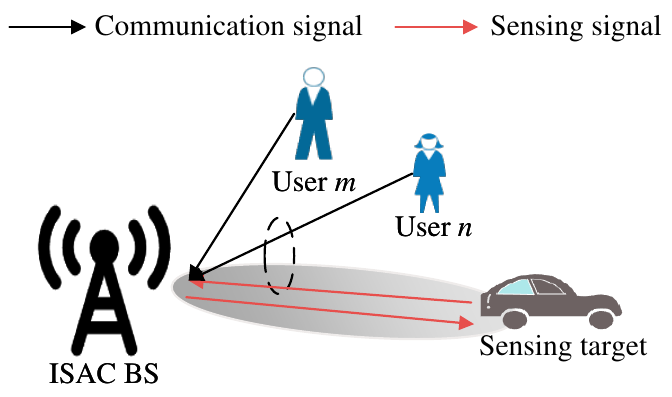}
	\caption{ llustration of uplink MI. } 
	\label{uplink}
\end{figure}
As depicted in Fig. \ref{uplink}, communication uplink signals and sensing echo signals mutually interfere in the integrated receiver.

\subsubsection{Interference Suppression}
A straightforward method to mitigate MI between uplink communication and sensing echo signals is to design longer signal transmission intervals, ensuring that radar echo signals always arrive before uplink communication signal transmissions. 
However, when sensing targets are close, excessively long transmission intervals can sharply decrease communication transmission rates. 
D’Andrea \textit{et al.} minimized MI at the communication receiver and suppressed the impact of sensing on communication by employing uplink channel estimation algorithms and receiver data detection algorithms \cite{8871348}. 
Building on this, \cite{9618653} proposes a method to separate and detect uplink signals and sensing echoes using channel diversity between UEs and targets. Specifically, communication signals are first detected and demodulated, and then subtracted from the received signal to obtain the sensing echo signals.

In NOMA-assisted uplink ISAC system, interference between communication signals and radar echo signals can be effectively eliminated by designing an SIC framework.
For instance, in the signal processing algorithm devised by Ouyang \textit{et al.}, the BS initially treats radar waveforms as interference to decode communication signals, then subtracts the communication signals from the received signal, leaving the remainder for radar sensing \cite{9782407}. 
Furthermore, in \cite{10163897}, the authors investigate two SIC sequences: communication-centric SIC (C-SIC) and sensing-centric SIC (S-SIC). 
In C-SIC, communication signals are processed first, followed by radar echo signals, while the order is reversed in S-SIC. 
These works confirm that combining NOMA and SIC technologies can offer higher degrees of freedom (DoF) than frequency-division ISAC system at the same communication rate.

\subsubsection{Interference Avoidance}
Chen \textit{et al.} proposed a sensing-aided Kalman filtering (KF)-based CSI estimation method for UL ISAC in \cite{9990573}.
Specifically, the first step involves utilizing the multiple signal classification (MUSIC) algorithm to obtain users' AOA. Then, this information is used as prior knowledge to enhance the estimation of uplink CSI, thereby improving communication reliability. 
However, the received signals in the abovementioned studies consist only of direct and scattered rays from user-transmitted signals, without considering echo signals in the downlink sensing. As a result, sensing needs more effective control methods, making continuous environmental sensing challenging to achieve. 
To address this issue, the authors propose an integrated framework for concurrent downlink and uplink in \cite{10036139}. To mitigate the impact of uplink communication on sensing echo signals, the SIC method is employed at the receiver to enhance the received echo signals. 

Additionally, flexible beam management is also necessary to avoid uplink MI. In this regard, Wang \textit{et al.} divided sensing into two stages: wide-beam detection and fine-beam parameter estimation \cite{9645576}. In both stages, the impact of uplink sensing on communication is eliminated by maximizing the communication SINR. In \cite{9171304}, a sensing-assisted beam alignment algorithm based on the extended KF framework is proposed to reduce the impact of sensing on communication. This method reduces the overhead of beam tracking in high-speed mobile scenarios and does not require additional sensing devices.

\subsubsection{Interference Exploitation}
The uplink provides a priori information to the downlink CSI acquisition process, including the pilot design, channel estimation, and feedback, thus assisting in reducing the feedback overhead. \cite{9570376}.  
However, due to the short coherent intervals, obtaining instantaneous values of CSI information becomes challenging. Therefore, Subhash \textit{et al.} proposed a statistical channel estimation algorithm assisted by RIS, which achieves over 100\% gain in fair SNR allocation among different users, with only 40 RIS elements \cite{10167745}. 
Furthermore, You \textit{et al.} proposed a strategy for designing uplink transmission based on partial CSI, including statistical CSI between the RIS and users and instantaneous CSI between the RIS and the BS \cite{9309152}. 
Under this framework, joint optimization of user-side transmission precoding and RIS beamforming is conducted to eliminate the impact on sensing. 
To relieve the high training overhead of ISAC channel estimation in highly mobile scenarios, Liu \textit{et al.} proposed a high-mobility downlink channel estimation scheme assisted by the uplink \cite{9110823}. 
Specifically, they utilize time-domain signal sampling from the uplink to recover channel parameters of individual physical scattering paths, including angle, delay, Doppler frequency shift, and channel gain. Based on this information, a delay-Doppler-angle domain large-scale MIMO downlink channel estimation can be constructed. 

\section{Single BS Clutter Management}
The stationary clutter signals reflected from permanent or long-period static objects, characterized by unchanged characteristics over time and near-zero Doppler frequencies are considered to carry limited useful information and thus treated as stationary clutter. 
However, when the BS operates in complex geographical, meteorological, and electromagnetic environments, it receives non-uniform clutter signals and requires more independently and identically distributed training samples for the estimation of clutter covariance matrix (CCM). 
Additionally, rich multipath environments in mobile networks introduce non-stationary clutter, which poses significant challenges to sensing accuracy. 
This section overviews from clutter suppression, clutter avoidance and clutter exploitation. A summary of clutter management methods is provided in Table
\ref{clutter}.
\setlength{\extrarowheight}{4pt}
\begin{table*}[htbp]
	\centering
	\caption { Summary of Clutter Management Methods}
	\label{clutter}
	\begin{tabular}{|l|p{3.5cm}|p{1.5cm}|p{8cm}|}
		\hline
		\textbf{Scopes} & \textbf{Specific Methods}  
		& \textbf{References} & \textbf{Summary in one sentence} \\ \hline			
		\multirow {2} {*} {Clutter suppression} 
		& Physical Methods  &  \cite{lazaro2014techniques}   &  Increase the height of BS installation and install clutter isolation devices around BS.    \\ \cline{2-4}
		& Filter Design & \cite{WOS:000082958500009, WOS:000391826300005, warde2017staggered, 8827589, luo2023integrated, 7792596, 8289376, 8960304,9223705, han2016novel,zhang2019reduced}   & 
		Identify and eliminate clutter through time-domain, frequency-domain, and space-time domain approaches. \\ \hline 
		
		\multirow {2} {*} {Clutter Avoidance}  
		&  Waveform Design &  \cite{li2017joint, liu2022joint, tang2016joint, 7421368, wang2017robust}  & Optimize waveforms based on clutter prior information. \\ \cline{2-4} 
		& Flexible Antenna Structures & \cite{liu2022joint, 9508883, 9998527, 10149583, 10141975, 9893567, 9696209,wei2023ris} & Adjust the phase and amplitude of the transmitted signal to effectively suppress clutter at the receiver. \\ \hline
		
		\multirow {1} {*} {Clutter Exploitation} 
		&Clutter Detection and Signal Processing & \cite{9764308, gao2023uav,cooper2021autonomous,fung2020using,9337372}   & 
		Environmental clutter plays a significant role in environmental monitoring, environmental reconstruction, and emergency rescue. \\ \hline
		
	\end{tabular}
\end{table*}

\subsection{Clutter Suppression}
\subsubsection{Physical Methods}
Firstly, the ISAC BSs with higher height can increase the tilt angle of array to increase the likelihood of direct-path propagation.
However, the costs also rise with the increase of BS height, making it impractical for large-scale deployment in urban areas. 
Secondly, installing clutter-rejecting structures around the BS can visually prevent clutter. 
However, these structures may weaken the reception of useful signals while blocking clutter reception. 
Although these methods can somewhat alleviate clutter, it is evident that due to the unpredictable statistical characteristics and close coupling with useful signals, relying solely on physical elimination methods is challenging to meet high sensing demands \cite{lazaro2014techniques}.

\subsubsection{Filter Design}

From different signal processing perspectives, filter design can be categorized into time-domain filters, frequency-domain filters, and space-time adaptive processing. Their main characteristics are as follows.

\begin{itemize}
	\item \textbf{Time-Domain Filters: } 
Time-domain regression filters are specifically designed to remove stationary clutter centered around 0 ms$^{-1}$. However, The SNR of valuable echoe signals decreases when they arrive within the filter duration \cite{WOS:000082958500009}.  
Therefore, it is crucial to design appropriate clutter detection algorithms before clutter removal. For instance, clutter environment analysis using adaptive processing can dynamically adjust clutter suppression levels based on the amount of ground clutter in the received signal, which has been proven feasible both in sequential \cite{WOS:000391826300005} and staggered pulse repetition intervals \cite{warde2017staggered}. 
Assuming that sensing parameters remain constant within the coherence period, each path in the received signal will only exhibit phase differences caused by Doppler frequency shift at different times. Based on this, Rahman \textit{et al.} proposed a background clutter removal method using straightforward recursive calculations. 
Specifically, the clutter estimation is updated every second from the estimations of neighboring channel matrices. Once a stable estimation is obtained, it is subtracted from current and future channel estimations within subsequent time intervals to improve channel estimation accuracy \cite{8827589}.

\item \textbf{Frequency-Domain Filters: }
With differences in the center frequency of clutter and moving targets in the power spectrum, moving target indication (MTI) and moving target detection (MTD) methods can filter out clutter while retaining targets. MTI processing is mainly used to suppress clutter from stationary objects, with pulse cancellers initially employed in engineering to suppress clutter at zero frequency \cite{9032372}. 
Adaptive MTI filters are often used to suppress clutter with non-zero average Doppler frequencies by dynamically adjusting the filter's null position.
MTD consists of a set of bandpass filters with deep nulls at zero frequency. Building on clutter suppression, MTD can provide velocity information of moving targets, enabling the differentiation of targets from clutter background \cite{8506724}. 
Luo \textit{et al.} observed that the equivalent echo channel vector caused by static environments remains unchanged across multiple OFDM symbols, thus they proposed a novel approach called mean phasor cancellation (MPC). This approach involves constructing a hybrid sensing channel model that encompasses both static environments and dynamic targets. By employing angle-Doppler spectrum estimation during the echo arrival phase, static clutter can be filtered out, facilitating dynamic target detection and angle estimation \cite{luo2023integrated}. 

\item \textbf{Space-Time Adaptive Processing (STAP):  } 
The key to STAP lies in the accurate estimation of the CCM, which reflects the clutter characteristics of the cell under test (CUT) \cite{7792596}.
Sun \textit{et al.} emphasized the method of accurately estimating the CCM using the training data from range gates close to the CUT to maximize the sensing signal-to-clutter-plus-noise-ratio (SCNR) \cite{8289376}.
Hu \textit{et al.} revealed the cyclical nature of the CCM and, leveraging this insight, devised a novel STAP algorithm. This approach significantly enhances the accuracy of CCM estimation by integrating cyclical CCM constructs across time, space, and space-time dimensions alongside conventional STAP techniques \cite{8960304}. 
In the scenario of limited training data, Sun \textit{et al.} developed a low-rank estimation algorithm based on the Kronecker product structure to estimate the CCM. This approach not only simplifies the training process but also enables effective target detection through a single channel \cite{9223705}. 
Moreover, Han \textit{et al.} proposed an approach that approximated clutter positions using training samples, which allowed for the recovery of clutter amplitudes and the estimation of CCM. This method effectively reduced the dimensionality of sparse recovery problems, consequently enhancing the efficiency of STAP \cite{han2016novel,zhang2019reduced}.
\end{itemize}

	\subsection{Clutter Avoidance}
	\subsubsection{ Waveform Design}
Li \textit{et al.} utilized the low-rank properties of sparse sensing radar and devise a waveform scheme employing random unitary waveform matrices. This design preserves radar waveform flexibility while effectively achieving spectrum sharing between radar and communication systems in cluttered environments through centralized joint optimization. It enhances radar SINR and communication rates while ensuring power efficiency \cite{li2017joint}.
However, the above waveform optimization based on spatio-temporal two-dimensional statistics makes it difficult to obtain precise sensing results. Therefore, Liu \textit{et al.} introduced a symbol-level precoding scheme by integrating the STAP algorithm to optimize the spatio-temporal transmission waveform for each transmitted symbol \cite{liu2022joint}.
Due to the slow speed of moving targets, their Doppler frequency shift is relatively small, making them more susceptible to environmental clutter interference.
To address this issue, Tang \textit{et al.} proposed a waveform design scheme that prioritizes mitigating clutter interference from neighboring target distance units to reduce the potential impact of strong clutter. Compared to linear frequency modulation (LFM) and OFDM waveforms, the proposed algorithm exhibits the higher SINR at low Doppler frequencies \cite{tang2016joint}. 

However, the effectiveness of clutter suppression in the aforementioned study relies on prior information about the targets and clutter. 
When environmental parameters are inaccurate, or there is a significant frequency offset in the received signals, target detection and estimation performance will deteriorate. 
Chiriyath \textit{et al.} evaluated the influence of clutter on S\&C dual functionalities in \cite{7421368}. 
The results indicate that clutter caused by moving objects significantly lowers the performance boundary. The CCM induced by moving objects contains errors. 
Therefore, Wang \textit{et al.} proposed a robust joint design strategy to address the uncertainty in the target steering vector. 
This strategy optimizes MIMO radar transmission and reception filters to maximize SINR, thereby effectively suppressing clutter and enhancing the robustness of target detection \cite{wang2017robust}.
	
\subsubsection{ Flexible Antenna Structures}
	
RIS-assisted clutter avoidance involves altering the phase and amplitude of transmitted signals to illuminate targets effectively. The reconfigurable holographic surface (RHS) serves as a metasurface with an embedded feeding network deployed on transceivers to control radiation amplitude better. This section reviews clutter suppression methods encompassing passive RIS, active RIS, and RHS.

\begin{itemize}
	\item \textbf{Passive RIS: }
	The inclusion of RIS can help avoid clutter by altering its direction, even in scenarios with weak or blocked direct channels. For instance, Liu \textit{et al.} deployed multiple RISs in a MIMO radar system to achieve better clutter avoidance and multi-target detection performance \cite{liu2022joint}. 
	In NLoS scenarios, changing RIS parameters regarding beamforming (size and direction) and polarization of electromagnetic waves can establish favorable artificial propagation/sensing scenarios \cite{9508883}. 
	However, systems involving passive RIS inevitably face the challenge of the ``multiplicative fading'' effect. Hence, the capacity gain obtained in strong LoS link scenarios is limited \cite{9998527}.
	
	\item \textbf{Active RIS: }
	Active RIS can amplify reflected signals by integrating amplifiers into their components. 
	However, this comes at the cost of significantly higher complexity, noise, and power consumption compared to passive RIS. 
	Thus, in \cite{10149583}, Wang \textit{at al.} developed a hybrid RIS mode that can configure the phase and modulus of the incident signal by absorbing part of the signal energy. 
	The results showed that this approach can fulfill target detection functions at lower transmission power than active RIS alone while achieving higher communication transmission rates than passive RIS alone.
	Additionally, Liao \textit{at al.} designed an active RIS-assisted ISAC scheme to achieve simultaneous multi-target sensing in NLoS areas and communication with multiple users \cite{10141975}. 
	To avoid clutter, interference power constraints for each clutter are introduced while maximizing the minimum sensing beam gain between multiple targets. 
	Notably, under the assumption of perfect communication/sensing CSI, active RISs can enhance sensing performance, especially in NLoS areas.
	
	\item \textbf{RHS: }
	RHS is a new type of antenna with simple hardware and compact unit spacing. Unlike traditional phased array beamforming schemes, RHS generates beams using an amplitude-controlled structure. Therefore, holographic radar consumes less power for the same detection accuracy and cost than phased array radar \cite{9893567}.
	However, beamforming needs to be optimized at the BS and RHS, and then combined at the communicating receiver and sensing receiver, delivering additional interference suppression gains while increasing signal processing complexity. \cite{9696209}. 
	Combining the solid focusing ability of RHS with the ability of RIS to compensate for NLoS links, Wei \textit{et al.} proposed a novel broadband DFRC system design to reduce the impact of environmental clutter on target sensing, where the transceiver is equipped with RHS and the RIS reflector is deployed in the channel \cite{wei2023ris}.
	
\end{itemize}

\subsection{Clutter Exploitation}

Electromagnetic waves exhibit abundant scattering, diffraction, and refraction phenomena during spatial propagation, providing rich multipath signals. These clutter signals play a crucial role as useful signals in various applications such as environmental monitoring, environmental reconstruction, and emergency rescue, which will be introduced in the following separately.
\begin{itemize}
	\item \textbf{Environmental Monitoring:} The Doppler weather surveillance radar network, known as next-generation radar (NEXRAD), consists of 159 high-resolution S-band polarimetric Doppler weather radars. The primary function of NEXRAD is to detect and track precipitation, assisting in predicting atmospheric threats to life and property \cite{9764308}.

\item \textbf{Environmental Reconstruction:}
Gao \textit{et al.} proposed a simultaneous localization and mapping (SLAM) framework for the positioning and 3D environmental mapping of unmanned aerial vehicles (UAVs) \cite{gao2023uav}. 
 As executors of detection and data collection tasks, UAVs can autonomously collect data and reconstruct 3D scenes in hazardous environments such as pipelines or boilers, based on received environmental clutter. 
Additionally, Cooper \textit{et al.} represented clutter fields using the spatial density and permeability at each coordinate \cite{cooper2021autonomous}. 
Approximating clutter fields allows clutter profiles to become simple geometric objects. Thus, the clutter profiles corresponding to vehicle clutter thresholds can serve as obstacle boundaries for obstacle avoidance.

\item \textbf{Emergency Rescue: } 
Fung \textit{et al.} proposed a deep learning-based urban search and rescue (USAR) model, where mobile robots collect datasets in the environment to identify victims based on differences in occlusion and illumination variations \cite{fung2020using}. 
Additionally, Heuermann \textit{et al.} introduced a maritime search and rescue system based on S-band illuminating harmonic radar. By collecting echo signals and classifying them, life jackets and rescue boats were tagged, and tests were conducted in the Baltic Sea \cite{9337372}.
	
\end{itemize}

\section{CLI Management in CoMP-ISAC System}	
\begin{figure*}
	\centering
	\includegraphics[width=0.8\linewidth]{ 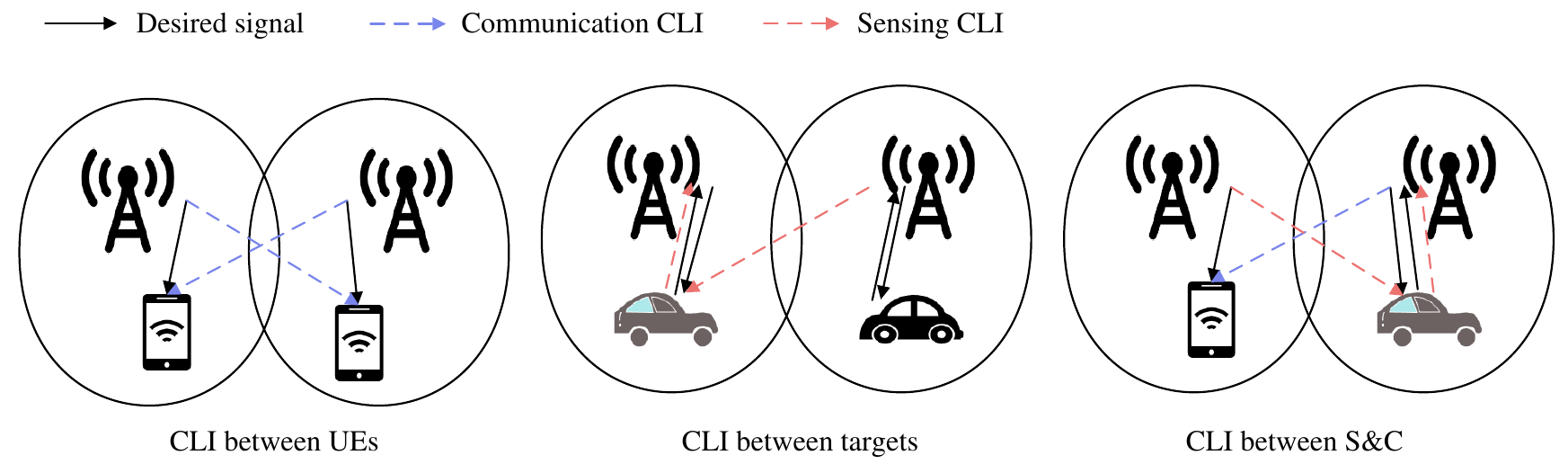}
	\caption{ llustration of CLI in multi-ISAC systems \cite{9139384}. } 
	\label{cross link}
\end{figure*}
In networks composed of multiple ISAC systems, the complexity of information interaction increases significantly with the sharp rise of S\&C links, leading to a growing concern over CLI. 
Specifically, the CLI can be categorized into interference between communication links, interference between sensing links, and crosstalk between sensing and communication links, as shown in Fig. \ref{cross link}.
The crosstalk in CoMP-ISAC systems originates from multiple BSs.
However, in cluster networks based on CoMP based ISAC systems can integrate the communication and sensing capabilities of multiple BSs, facilitating a wide-area coverage and ubiquitous sensing \cite{9139384, 9180098}. 
To eliminate CLI in CoMP-ISAC system, this section provides an overview from the perspective of the cooperative sensing approach, uplink and downlink cooperation, and network resource allocation.

\subsection{Collaborative Sensing Approach}

The operation modes of BS can be divided into uplink reception mode and downlink transmission mode. 
In CoMP-ISAC systems, coordinating the operation modes of multiple BSs can fully utilize the macro-diversity gain of distributed BSs and suppress the generation of interference \cite{10308585}. Specifically, the BS cooperation modes mainly include self-transmit self-receive, multi-transmit self-receive, and multi-transmit multi-receive modes. 

\subsubsection{Self-Transmit Self-Receive}
Chen \textit{et al.} proposed a strategy to optimize communication and sensing performance in multi-BS systems through collaborative waveform design and clustering.  
This method allows BSs to automatically join the communication and sensing clusters when performing communication and sensing tasks, respectively. \cite{10015103}. 
Furthermore, ensuring that all BSs in the cluster operate in the same link direction (uplink or downlink) can effectively prevent strong CLI between neighboring BSs. 
For instance, Sun \textit{et al.} proposed a clustering scheme using mutual coupling loss (MCL) as a metric, where pairs of BSs with MCL values smaller than a certain threshold are clustered together to prevent inter-BS interference \cite{7390863,7565057}.

\subsubsection{Multi-Transmit Self-Receive}

Xu \textit{et al.} introduced a collaborative transmission and reception scheme where one BS is selected as a receiver. In contrast, the remaining stations transmit dedicated sensing waveforms and communication signals \cite{xu2023integrated}. 
To suppress MUI and MI, a joint optimization method for the transmission and reception beamforming matrix is proposed, aiming to minimize the adaptation error of the sensing beam direction.

However, this collaborative approach may lead to insufficient transmission power when all communication users are served by all BSs.
To address this challenge, Xu \textit{et al.} proposed a dynamic sensing receiver selection method to enhance collaborative sensing gain \cite{xu2023joint}.
Compared to a fixed sensing reception mode, this approach can reduce the transmission power of the BSs and improve the detection probability. 
Furthermore, the selection of communication users is also based on the principle of minimizing transmission power to suppress inter-user interference more effectively.

\subsubsection{Multi-Transmit Multi-Receive}
In \cite{9842350}, Huang \textit{et al.} devised a cooperative mode among BSs, where each ISAC transmitter sends communication signals to a UE. 
After being reflected by targets, the echo signals are collected by sensing receivers to estimate target positions. 
By coordinating and controlling the transmission power of each BS, they aim to reduce inter-BS interference while lowering the overall system's transmission power. 
However, when targets are located between two BSs, they experience direct paths interference, leading to a sharp decline in sensing performance. 
To flexibly balance the performance between S\&C and avoid MI, Li \textit{et al.} introduced a functionality selection module. 
They proposed a signal processing algorithm based on a redundant fusion of multi-view data to fully utilize the multi-directional sensing information obtained from distributed sensing nodes \cite{9916293}. 
Additionally, the critical contribution lies in deriving a fusion model through hypothesis testing and optimal voting analysis, providing a basis for predicting fusion accuracy.

Furthermore, with the innovation of cell structures, attention has been paid to joint BS mode selection, transmission beamforming, and receiver filter design in cell-free cooperative ISAC networks. 
For instance, Liu \textit{et al.} proposed a system design scheme that maximizes the sum of the sensing SINR while meeting the requirements of communication QoS, total power budget, and constraints on the number of transceivers \cite{unknown}. 
In cell-free systems, signal transmission and reception have greater DoF, enabling a more flexible collaboration for S\&C and various interference avoidance. 
However, this puts forward challenges in resource scheduling and information processing, encouraging the development of low-complexity sensing algorithms and receiver interference cancellation algorithms.

\subsection{Cooperation of Uplink and Downlink }

In mobile sensing networks, the complex communication links and sensing echo signals between multiple BSs lead to sensing ambiguity and MI between S\&C due to spatially separated asynchronous transceivers, limiting the improvement in sensing accuracy. 
To tackle this challenge, Ni \textit{et al.} proposed an uplink-aided sensing scheme \cite{9349171}. 
On the one hand, they utilized cross-antenna cross-correlation operations to mitigate the timing offset (TOs) and carrier frequency offset (CFO) issues between the asynchronous sensing receiver and transmitter. 
On the other hand, a high-precision AoA estimation algorithm that jointly processes uplink signals in the time-space and frequency domains is proposed to improve the estimation accuracy compared to using spatial samples alone. 
Moreover, D’Andrea \textit{et al.} designed various uplink channel estimation and data detection receivers, ranging from simple channel-matched beamformers to zero-forcing and linear minimum mean square error receivers, to suppress interference from environmental scatters \cite{8871348}. 
Besides, integrated systems with uplink-downlink collaboration achieve higher sensing rates while maintaining the same communication rate \cite{9782407, 9800940}.

\subsection{Network Resource Allocation}
In CoMP-ISAC systems, power control between sensing and communication signals directly affects the interference between S\&C, inter-user interference, and inter-BS interference. 
Although reducing the transmission power reduces interference, it also weakens the desired signal.
Moreover, increasing transmission power does not improve the overall EE. 
Therefore, network resource allocation among  multiple BSs becomes crucial, aiming to maximize desired signal enhancement with minimal power usage.

\subsubsection{Network Architecture Design }
Network architecture design covers the exploration of emerging network elements, network configuration, and routing protocol.
\begin{itemize}
	\item \textbf{Emerging Network Elements: } 	
One network architecture involves integrating sensing capabilities into a C-RAN. 
For instance, some dedicated remote radio units (RRUs) can be configured as specialized sensing receivers during the downlink sensing \cite{9296833}. 
Another network structure integrates passive target monitoring terminals (TMTs), such as radar and cameras, into traditional cellular networks \cite{9802828,9143269}. This setup makes the BS act as the transmitter while the TMTs serve as the receivers, thereby avoiding self-interference issues.

\item \textbf{Network Configuration: }
ISAC-assisted neighbor discovery is an important step in network initialization that can provide rich prior information for medium access control (MAC) protocols, routing protocols, and topology control protocols. 
Therefore, sensing information can expedite traditional neighbor discovery. 
For example, a directional discovery algorithm based on a scanning network radar sensing algorithm is proposed in \cite{7023734}, which utilizes the approximate positions of neighbors obtained through sensing to accelerate neighbor discovery and avoid CLI.
Moreover, Wei \textit{et al.} combined ISAC and the Gossip protocol to develop a dynamic neighbor discovery algorithm based on sensing data while reducing the communication resource consumption, thus providing a new network configuration in CoMP-ISAC system\cite{10251148}.

\item \textbf{Routing Protocol: }
In scenarios with high demand for network throughput, whether static or dynamic, network congestion is a primary cause of decreased throughput and increased latency. 
Narayan \textit{et al.} proposed a routing protocol that monitors and reports the average queue utilization threshold of multiple interfaces as a QoS parameter \cite{6450550}. 
During congestion, this protocol balances the load across multiple paths, thereby enhancing network throughput and effectively avoiding packet collisions.
\end{itemize}

\subsubsection{Power Allocation}
The water-filling power allocation algorithm is a common method used for antenna joint deployment and coordination. 
More transmission power is allocated to BSs with better channels.
However, allocating high power alone does not prevent energy consumption and interference. 
Therefore, Zhao \textit{et al.} combined the linear precoding algorithm with the water-filling power algorithm to address the interference effects between users \cite{8367178}.

Optimizing the power allocation between BSs while simultaneously minimizing the overall transmission power is essential for enhancing the system's EE. 
Thus, Li \textit{et al.} jointly optimized beamforming and receiver filter design in a cooperative dual-cell network. By optimizing the transmission and reception processes between two BSs, they reduced the total transmission power of the system \cite{9835135}.
Additionally, Liu \textit{et al.} demonstrated that the application of CoMP combined with SIC methods in a multi-user NOMA cluster system not only eliminates interference but also achieves higher system throughput, highlighting the applicability of collaborative interference cancellation \cite{8861122}.

Moreover, it is essential to consider the adjustment of BS power budgets and the collaborative power management between UEs to optimize the network power control strategies \cite{Haili2013PowerAS}. 
Specifically, BSs first need to sense the electromagnetic environment and establish models for interference channels and S\&C channels \cite{9945983}. 
Furthermore, BSs can intelligently adjust the transmission power of UEs in the uplink time slots and their own transmission power in the downlink time slots to effectively suppress interference. 
This bidirectional power control strategy not only improves spectrum efficiency but also enhances the overall performance of the network and user QoS.

\section{Future Trends}
\begin{figure}
	\centering
	\includegraphics[width=1\linewidth]{ 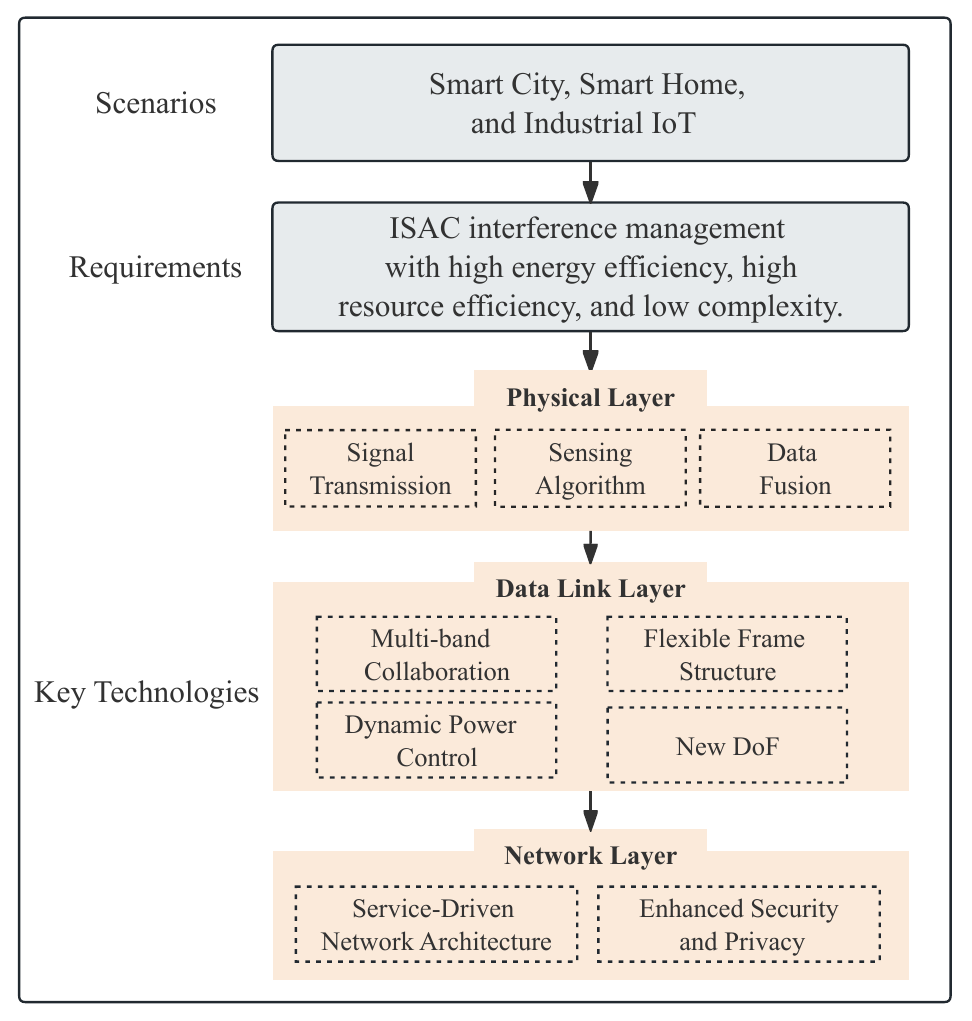}
	\caption{ Future research trends of ISAC interference management. } 
	\label{future trends}
\end{figure}
As shown in Fig. \ref{scene}, the emerging applications will be carried by intelligent interconnected 6G networks. 
In this section, we summarize the potential interactions and connections between ISAC and other emerging communication technologies, aiming for high spectrum efficiency, high EE, and low complexity algorithm. 
The insights of future research trends of ISAC interference management are summarized in Fig. \ref{future trends}.
\subsection{Signal Processing}

To accommodate intelligent application scenarios, ISAC signal processing imposes higher requirements on sensing accuracy and algorithm complexity. 
Therefore, the signal transmission, sensing algorithm, and data fusion are crucial.

\subsubsection{Signal Transmission}

Two challenging research directions are provided, namely advanced transceiver design and cooperation between uplink and downlink for additional interference suppression gain and efficient utilization of resources. 

\begin{itemize}
	\item \textbf{Advanced Transceiver Design: }	
Interference avoidance and interference mitigation are primarily handled by the transmitter and receiver, respectively. 
The matching design between transceivers is important to suppress interference and improve signal processing efficiency.
However, most existing research focuses on waveform optimization based on existing waveforms and alternate optimization-based transceiver beamforming design, overlooking the differences in new waveforms and transceiver channels. 
In response to high data rate demands, mmWave signals and frequency-hopping signals provide extended solutions for transceiver design. 
Moreover, carrier aggregation can further improve spectrum resource utilization efficiency \cite{10285442}. 
Additionally, modulation techniques like index modulation offer new DoF for signal transmission, proving an effective method for meeting high throughput requirements. In challenging indoor scenarios with severely limited signal transmission ranges, traditional far-field channel modeling struggles to meet the high-precision sensing requirements. Considering the inevitable trend of future communications expanding into higher frequency bands, transceiver designs based on more precise near-field channel modeling can provide more applicable interference suppression solutions.

\item \textbf{Cooperation Between Uplink and Downlink: } 
In uplink communication signal sensing, the flexibility and mobility of UEs offer additional DoFs, which are beneficial for enhancing spectrum resource utilization efficiency and the effectiveness of real-time sensing. 
However, the coupling between uplink signals and echo signals leads to MI. 
Therefore, uplink-downlink cooperation becomes crucial, aiming to fully utilize uplink communication signals to assist in reconstructing the physical and electromagnetic environment \cite{10036139}.
However, existing researches are commonly based on the assumption of channel orthogonality.
When sensing targets or environments change rapidly, the coherence time of the channel environment changes rapidly, making it challenging to meet the assumption of channel orthogonality. 
Therefore, future development of uplink-downlink cooperation should focus on researching terminal-assisted sensing algorithms in high-speed scenarios. 
Additionally, there is a need for a deeper exploration of the practical aspects of channel orthogonality to establish more accurate models for uplink-downlink cooperation.

\end{itemize}

\subsubsection{Sensing Algorithm}

Enhancing robustness and reducing complexity are dual objectives in refining sensing algorithms, aiming to improve their reliability under diverse conditions and efficiency in resource-constrained environments.
\begin{itemize}
	\item  \textbf{Enhancing Robustness: } 
Considering the complexity of the scenario, there is insufficient research on the impact of channel estimation errors, TO, CFO, and mobility estimation on sensing accuracy, as well as on the development of elimination algorithms for these factors. 
In imperfect self-interference cancellation or non-full-duplex radio scenarios, the noise floor of radar receivers will significantly increase, making it difficult to locate distant targets accurately \cite{10400873}. 
Additionally, with the development of high-frequency signals, the probability of channel estimation errors is high, posing inevitable challenges for detecting and eliminating interference signals. 
Therefore, interference cancellation under imperfect conditions is a worthwhile research problem. Future studies will focus on improving the accuracy and robustness of sensing algorithms to address various interference and error issues that arise in complex scenarios.

\item \textbf{Low Complexity: } 
For resource-constrained scenarios such as embedded systems, mobile devices, or low-power applications, managing interference within limited time, resources, and processor cycles should aim to minimize the complexity of designed sensing algorithms. 
However, traditional signal processing involves the sequential steps of sensing, communication, and computation, which can be inefficient \cite{10149589}. 
Future algorithms should consider time efficiency, space efficiency, computational efficiency, scalability, and hardware-friendliness comprehensively rather than solely focusing on reducing the power consumption of sensing signals.
Edge AI represents an intelligent signal processing approach encompassing edge learning and edge inference. 
It holds significant potential in the highly coupled and mutually reinforcing processes of data collection (sensing), information extraction (computation), and information transmission (communication). 
With its inherent learning and predictive capabilities, AI is expected to play a more significant role in resource scheduling, adaptive networking, and proactive prevention of network congestion and security threats \cite{10217150,wen2023integrated}.
	
\end{itemize}

\subsubsection{Data Fusion}
Single BS has limitations in coverage range and sensing angles. With the advancement of the networked mobile communication systems, multi-BS collaborative sensing emerges as the inevitable choice to meet the requirements for remote and precise sensing \cite{10273396}.
However, existing researches show that there still remain scopes for further enhancement in sensing accuracy with data-level fusion, while signal-level fusion demands excessively high synchronization accuracy \cite{Li2020ATO, 9534484}.
To tackle this problem, Wei \textit{et al.} proposed a symbol-level fusion method based on communication signals for sensing, enabling the estimation of target positions and velocities where the existing communication systems can meet the requirements for time and phase synchronization \cite{10226276}. 
While ``soft fusion'' algorithms directly weighting received signals achieve optimal accuracy, ``neutral fusion'' based on the fusion of sensing measurements and ``hard fusion'' based on the direct fusion of sensing results can effectively reduce the computational burden.
Therefore, the selection of future multidimensional information fusion methods needs to consider various factors, such as the signal processing capabilities of servers and receiving nodes, and the accuracy requirements of sensing tasks. 
Moreover, time synchronization, frequency synchronization, and phase synchronization among distributed sensing nodes are crucial for enhancing sensing fusion accuracy.

\subsection{Resource Allocation}
\subsubsection{Multi-Band Cooperation}
Compared to the fifth-generation, 6G occupies more extensive frequency bands, including low frequency, mmWave, THz, and visible light. 
Due to the suitability for specific applications of different frequency bands, some comprehensive optimization solutions are required to mitigate interference. 
Low-frequency communication signals have strong penetration and coverage capabilities but are limited by inadequate sensing precision and time-frequency resolution. 
As the frequency increases, wavelengths become shorter and the available bandwidths get wider, providing advantages in fine-grained sensing \cite{serafino2021photonics}. 
Therefore, integrating the advantages and characteristics of different frequency bands in ISAC systems offers the potential to address interference challenges encountered in single-frequency bands.

\subsubsection{Flexible Frame Structure }
To enhance multi-user communication and multi-target sensing capability while reducing interference in CoMP-ISAC systems, it is essential to design signals with good auto-correlation and poor cross-correlation characteristics \cite{10049817}.
However, parameter configurations are relatively fixed in existing communication signal frame structures, making it challenging to meet different service requirements. 
In the current researches, the limited time-frequency resources allocated for sensing signals severely restrict resolution. 
Moreover, the randomness of specific sensing targets in communication signals leads to ambiguity and increases signal processing complexity. 
Therefore, it is urgent to design flexible frame structures that can be dynamically adjusted according to network conditions and user requirements \cite{10457036}.
For example, adjusting the length and configuration of subframes to accommodate different services and channel conditions. Furthermore, dynamic power control strategies can be implemented in frame structure design to adjust transmission power based on the proximity of users to BSs, thereby reducing near-far effects and interference from neighboring cells.

\subsubsection{Dynamic Power Control }
Effective interference elimination typically imposes strict requirements on node mobility and information exchange between nodes. 
By dynamic power control, coordination of inter-BS interference and MUI can be achieved. 
However, the fixed cellular structure suffers from drawbacks such as uneven resource allocation and inefficient resource utilization. Therefore, both the non-cellular network architecture and BS operational mode selection are essential research directions. In the case of non-cellular networks, the network can extend beyond cell boundaries, allowing users to communicate directly with the nearest nodes (such as satellites, drones, ground stations, or other devices), thereby reducing interference caused by resource sharing between edge users in traditional cellular networks and neighboring cell users. 
Meanwhile, flexible BS operational modes can coordinate BS switching/modes and S\&C operational modes to reduce MI and power loss \cite{10328645}.

\subsubsection{New DoF }

RIS and UAV provide novel DoFs for interference management, with RIS offering precise control over signals and UAV supplying agile LoS paths for improved network capabilities.
\begin{itemize}
	\item \textbf{ RIS:}
	RIS technology offers significant advantages in wireless communication, particularly in interference suppression. 
	By precisely controlling signal reflection and refraction, RIS technology can significantly reduce interference at a low cost while providing new pathways for signal transmission by deploying RIS at environmental monitoring equipment, which are crucial for realizing extensive sensing networks. 
	Active RIS can enhance the directness and efficiency of communication by creating virtual LoS links, thus playing a vital role in the ISAC systems \cite{zhu2023joint}.
	In addition, a lower matching error can be achieved by jointly optimizing active beamforming at ISAC BS, passive beamforming at RIS, power allocation, and time allocation to obtain the desired sensing beampattern \cite{sun2024joint}.
 
	\item \textbf{ UAV: }
	The UAV-assisted ISAC system is an important application due to the UAVs' fully controllable maneuverability, which also faces many challenges. 
	For example, UAVs have limitations in information processing capability, energy supply, and information interaction \cite{10077116}. 
	In this regard, existing solutions face challenges in meeting the S\&C functions in high-dynamic and resource-constrained UAV deployment scenarios \cite{9716042}. 
	Thus, one prospective future research trend is the use of UAVs as mobile airborne platforms to assist with target tracking and SLAM with lower power consumption. This can be achieved by utilizing the abundant LoS paths provided by the UAV to enhance the useful signal power.
	
\end{itemize}

\subsection{Network Architecture}
\subsubsection{Service-Driven Network Architecture}
Network elements with sensing capabilities face challenges in collaboration mode selection, sensing algorithm design, and sensing service billing.
On the one hand, future network architecture are expected to emphasize service-driven intelligent resource allocation and network configuration optimization, which are essential for effective interference management. 
This architecture can dynamically respond to real-time service demands, ensuring efficient operation of critical services in interference environments. 
By integrating advanced sensing technologies, such as sensing network elements, the network can monitor interference levels in real time and take appropriate actions, such as adjusting power distribution, changing frequency resources, or optimizing signal paths, to minimize the impact of interference on user QoS \cite{chekired20195g}. 
Furthermore, a service-driven network architecture also supports the transfer of control signaling to the user plane, alleviating the burden on the control plane and further enhancing the network's interference resilience and overall performance.

On the other hand, distributed software-defined networking (SDN) controllers dynamically adjust networks to adapt the changes in wireless environment, effectively managing interference. 
However, the complexity of wireless networks and the demand for computational resources in automated control are substantial, especially in environments with high-security requirements, posing challenges to integrating SDN features. 
In data-intensive scenarios such as indoor positioning and vehicular networks, traditional signal processing and channel models are no longer sufficient to handle multimodal and multi-angle information \cite{5483108}. 
Therefore, by combining SDN and AI, deeper data analysis and pattern recognition capabilities can be provided, helping networks more accurately predict and adapt to interference, achieving more efficient resource allocation and interference management strategies.

\subsubsection{Enhanced Security and Privacy }
Due to the openness of wireless propagation media, sensing operations are exposed to potential security risks and privacy concerns \cite{9336039}. 
Current research utilizes the powerful sensing capabilities of wireless sensing to track specific terminals and comprehensively perceive the surrounding environment, including the position, shape, size, movement trajectory, and speed direction of objects. 
Moreover, this technology holds tremendous potential for applications in health monitoring, enabling real-time monitoring of respiratory rates and heartbeats. 
Additionally, it can perceive sensitive targets in specific areas, such as military facilities or equipment. 
While these features bring new application possibilities for 6G systems, they also introduce new security challenges, particularly concerning unauthorized access and data leakage \cite{sun2023security}. 
Future 6G ISAC systems require the development of new architectures and protocols to enhance existing security mechanisms for wireless sensing services.
This includes but is not limited to enhanced encryption techniques, refined access control policies, rigorous monitoring of the data transmission process and directional modulation technology \cite{10045696, 10219052}.

\section{Conclusion}
This article presents a comprehensive survey on interference management methods of ISAC system, including SI, MI, clutter and CLI.
First, we introduce the ISAC channel modeling and performance metrics used in ISAC systems. 
Then, detailed reviews, analyses, and comparisons of interference management methods are provided from suppression, avoidance, and exploitation perspectives.  
Finally, we outline important challenges and future research directions. 
This paper can potentially serve as a reference for research on interference management within ISAC system.

 \bibliographystyle{IEEEtran}	
\bibliography{ref}	
\end{document}